\documentclass[review]{elsarticle}

\usepackage{lineno,hyperref}
\usepackage{caption}
\usepackage{color, soul}
\usepackage{subcaption}
\usepackage{changepage}
 \usepackage{wrapfig}
 \usepackage{threeparttable}
 \usepackage{dcolumn}
  \newcolumntype{d}{D{.}{.}{-1}}
 \usepackage{nomencl}
  \makeglossary
 \usepackage{cancel}
 \usepackage{fancyvrb}
  \fvset{fontsize=\footnotesize,xleftmargin=2em}
 \usepackage{lettrine}
 \usepackage{hyperref}
  \usepackage{graphicx, color, psfrag, float}
  \usepackage{pstricks,pst-node,pst-plot,pst-circ,moredefs}
  \usepackage{dcolumn}
\usepackage{bm}
\usepackage{epstopdf}
\usepackage{multirow}
\usepackage{booktabs}
\newcolumntype{L}{>{\centering\arraybackslash}m{3cm}}
\journal{Journal of Fluids and Structures}

\begin{document}

\begin{frontmatter}

\title{Vortex dynamics and Reynolds number effects of an oscillating hydrofoil in energy harvesting mode}

\author[mymainaddress]{Bernardo Luiz R. Ribeiro\corref{mycorrespondingauthor}}
\cortext[mycorrespondingauthor]{Corresponding author}
\ead{rocharibeiro@wisc.edu}

\author[mysecondaryaddress]{Sarah L. Frank}

\author[mymainaddress]{Jennifer A. Franck}

\address[mymainaddress]{Engineering Physics, University of Wisconsin-Madison, Madison, WI}
\address[mysecondaryaddress]{Mechanical Engineering, University of California, Berkeley, CA}

\begin{abstract}
The energy extraction and vortex dynamics from the sinusoidal heaving and pitching motion of an elliptical hydrofoil is explored through large-eddy simulations (LES) at a Reynolds number of $50,000$.  The LES is able to capture the time-dependent vortex shedding and dynamic stall properties of the foil as it undergoes high relative angles of attack. Results of the computations are validated against experimental flume data in terms of power extraction and leading edge vortex (LEV) position and trajectory. The kinematics for optimal efficiency are found in the range of heave amplitude $h_o/c=0.5-1$ and pitch amplitude $\theta_o=60^{\circ}-65^{\circ}$ for $fc/U_{\infty}=0.1$ and of $h_o/c=1-1.5$ and $\theta_o=75^{\circ}-85^{\circ}$ for $fc/U_{\infty}=0.15$. Direct comparison with low Reynolds number simulations and experiments demonstrate strong agreement in energy harvesting performance between Reynolds numbers of $1000$ to $50,000$, with the high Reynolds number flows demonstrating a moderate $0.8-6.7\%$ increase in power compared to the low Reynolds number flow. In the high Reynolds number flows, the coherent LEV, which is critical for high-efficiency energy conversion, forms earlier and is slightly stronger, resulting in more power extraction. After the LEV is shed from the foil, the LEV trajectory is demonstrated to be relatively independent of Reynolds number, but has a very strong nonlinear dependence with kinematics. It is shown that the LEV trajectories are highly influenced by the heave and pitch amplitudes as well as the oscillation frequency. This has strong implications for arrays of oscillating foils since the coherent LEVs can influence the energy extraction efficiency and performance of downstream foils.
\end{abstract}


\end{frontmatter}


\section{Introduction}

\begin{figure}[h]
\centering
\includegraphics[width=0.5\textwidth]{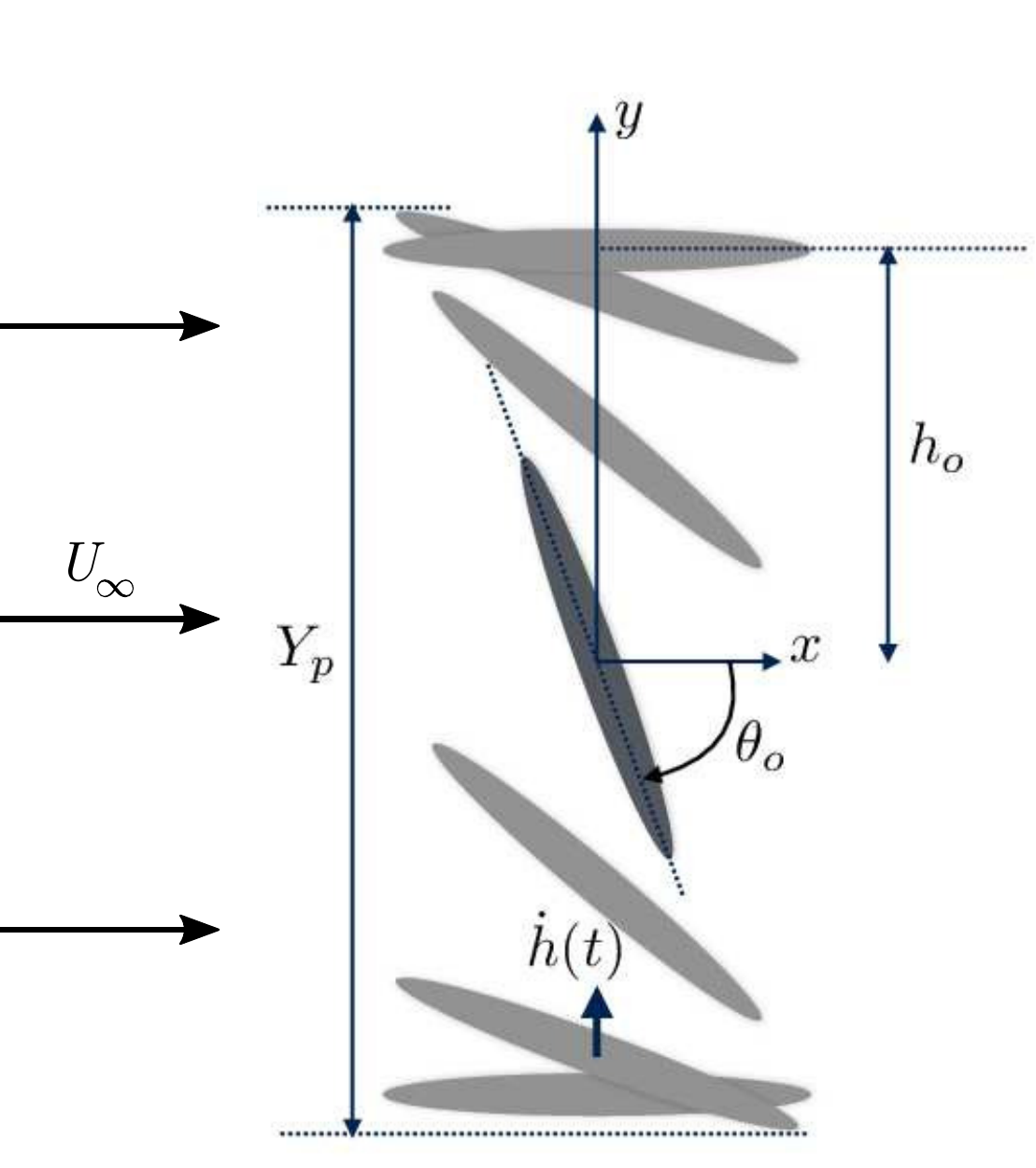}
\caption{Kinematics of the elliptical hydrofoil's motion shown for the upstroke.}
\label{f:stroke}
\end{figure}

Wave and tidal energy are estimated to hold $1420$ TWh/yr of extractable energy in the United States, about a third of the $4000$ TWh/yr of energy that is used in the United States \cite{doe}. Despite this rich source of clean, renewable energy there exist many engineering challenges in terms of the successful operation and maintenance of hydrokinetic turbines. Many devices designed for tidal energy extraction are in the form of rotational turbines, such as horizontal-axis turbines or Gorlov designs \citep{gorlov98}.  Another viable option is the use of an oscillating foil. To generate power during the upstroke, a foil heaves vertically with a positive angle of attack to produce a net positive lifting force and positive power.  It then repeats the symmetric stoke on the downstroke, with a pitch reversal at the top and bottom of the stroke (Fig. \ref{f:stroke}). Oscillating hydrofoils offer many advantages over rotational turbines, including avoiding the high tip speeds that scale with radius on rotating blades. The oscillating hydrofoil design can also fit in shallower waters than their rotating counterparts, and have the potential to be closely packed due to their simple geometry and more coherent wake structure. Another benefit is that the power extraction is largely based on the kinematics of the foil, such as operating frequency, pitch and heave amplitude.  Thus a single foil can very easily be optimized for power output over various flow speeds by modulating the kinematics without changing the size or overall design of the foil. The first modern mention of oscillating foils for energy harvesting applications was by McKinney and DeLaurier in 1981 \citep{mckdel}, but recently there has been renewed interest both experimentally and computationally in the form of various kinematic strokes, pitching axes, and flow conditions for optimal performance [\citealp{Young2014, Xiao2014}]. 

In the original exploration of oscillating hydrofoils, McKinney and DeLaurier defined the pitch and heave using sinusoidal functions, showing with both theory and experiments that the oscillating hydrofoil had the potential to perform at an efficiency comparable to rotating energy extraction devices \citep{mckdel}. Since then systems have been explored that completely drive the pitch and heave motion, or have at least one degree of freedom passively driven with a torsional or linear spring [\citealp{liu2013, SuThesis}]. In terms of completely driven systems, Simpson et al. performed experimental measurements on a NACA 0012 foil in a tow tank, characterizing the performance for a range of kinematic parameters in pure sinusoidal motion \citep{simpson2008}. An elliptic symmetric foil for easy tidal flow reversal was investigated experimentally by Kim et al. \citep{Kim2017} in a recirculating flume. Water flume confinement effects on energy harvesting performance have also been studied [\citealp{Yunxing2019, SuThesis}]. Various prototypes have also been designed and/or tested to date, including a $2$ kW prototype based on studies at Laval University \citep{kindum2011} and a $1$ kW prototype developed at Brown University [\citealp{cardona2015, MillerThesis}]. This manuscript, like many others in the literature, is concerned with the fluid mechanics surrounding the oscillating foil when it is in an energy harvesting kinematics regime, which limits the range of oscillation frequencies and amplitudes explored. Although the overall efficiency of an energy harvesting device will be lower than predicted values due to mechanical losses, the general trends in prototype data have been well predicted by simulations and scaled-down flume experiments [\citealp{cardona2015, MillerThesis}].

Kinsey and Dumas carefully explored a wide range of parameters computationally on a NACA 0012 foil, and found a peak efficiency of approximately 34\% \citep{kindum2008}. The optimal range of kinematics, also documented by other researchers [\citealp{zhu2011, wu2015}], has been found to be at a reduced frequency, $fc/U_{\infty}$, of $0.1-0.15$, where $f$ is frequency of oscillation, $c$ is chord length, and $U_{\infty}$ is the freestream velocity. The optimal phase difference between the pitch and heave cycles is approximately $90^{\circ}$, with heave amplitudes, $h_o/c$, within the range of $0.5-1$, and pitch amplitudes, $\theta_o$, within the range of $65^{\circ}$ to $75^{\circ}$. These optimal values have been demonstrated for a variety of slender foil shapes (including ellipses investigated in this paper) undergoing pure sinusoidal motion with pivot locations at or close to center chord. 

A key component to the optimal power production is the formation and timing of a coherent leading edge vortex (LEV) which enhances lift forces, and thus power, throughout the heave stroke [\citealp{Kim2017, Su2019}]. The LEV formation and shedding was shown to depend on the oscillating frequency and plunge amplitude of the foil by Baik et al. in their work on LEV dynamics and unsteady forces produced by pitching and plunging airfoils across different Reynolds numbers \citep{baik2012}. The large coherent vortices that are shed in the high-efficiency kinematic regime share a resemblance to bluff-body vortex-induced vibrations, which have also been applied towards hydrokinetic energy harvesting applications \citep{Bernitsas2014}. The trajectory of the highly coherent vortex structures are critical in placement of subsequent oscillating foils in array configurations \citep{Simeski2017}.

Another important parameter, especially in terms of informing the design of a large-scale prototype, is Reynolds number. Direct numerical simulations (DNS) at low Reynolds number on the order of $1000$ may not capture the complete flow physics of the experimental flume tests that are typically in the Reynolds number range of $30,000-50,000$ due to turbulent transition, or that of full-scale prototypes in a fully turbulent regime. A few groups have computationally explored higher Reynolds numbers, including Ashraf et al. \citep{Ashraf2011} and Xiao et al. \citep{Xiao2012} who both explored non-sinusoidal effects at moderate Reynolds numbers of $10,000-20,000$ using a two-dimensional Navier-Stokes solver. At much higher Reynolds numbers, Kinsey and Dumas \citep{kinsey2012} used an unsteady Reynolds Averaged Navier-Stokes (RANS) model with a Spalart-Allmaras turbulence closure for two-dimensional and three-dimensional hydrofoils to investigate tip effects and found good comparison with experimental results from a $2$ kW prototype with two-foils in a tandem configuration. Campobasso et al. \citep{campobasso2013} compared low ($Re=1100$) and high ($Re=1.5 \times 10^6$) Reynolds number results of a pitching and heaving foil using a compressible Navier-Stokes solver with a $k-\omega$ shear stress transport model, and found that the two regimes offer different dynamics in terms of optimal parameters for energy harvesting.

This paper continues to explore a sinusoidal heave and pitch stroke on an elliptical shaped foil, focusing on the effects of Reynolds number between $1000$, solved using two-dimensional DNS, and $50,000$, solved using a three-dimensional large-eddy simulation (LES). Direct comparison between these two Reynolds numbers will demonstrate the accuracy and limitations of low Reynolds number flow models compared with a regime an order of magnitude higher coinciding with laboratory flume experiments. The LES methodology allows for a resolved boundary layer which has been shown to accurately capture boundary layer separation and reattachment in unsteady flows \citep{franck2010}, compared to RANS models which often over-predict or do not fully capture unsteady vortex dynamics \citep{Greenblatt2004}. The three-dimensional LES will also capture spanwise fluctuations and momentum transport within the large-scale vortex structures that are impossible to discern with a two-dimensional DNS.

The computational results are compared against experimental data for validation by examining the forces and moments over the pitch/heave cycle, the efficiency of the stroke for optimal energy extraction, and the LEV formation and trajectory. Of particular interest in this manuscript is how the LEV and other large scale structures are formed and convected downstream by the kinematic motion of the foil, and the differences between the two distinct Reynolds number regimes.

\section{Numerical methods}

\subsection{Governing equations and numerical techniques}
To evaluate and compare power extraction capabilities of an oscillating hydrofoil in different kinematic modes, the finite-volume method is used to solve the incompressible Navier-Stokes equations in a non-inertial reference frame.  A two-dimensional DNS is performed at low Reynolds number ($Re=1000$), in which all the flow scales are fully resolved. For the higher Reynolds number simulations ($Re=50,000$) a wall-resolved LES is implemented with the spatially filtered incompressible Navier-Stokes equations,  

\begin{eqnarray}
\label{e:fNS}
\frac{\partial \bar{u}_i}{\partial t} + \bar{u}_j \frac{\partial \bar{u}_i}{\partial x_j} &=& -\frac{1}{\rho} \frac{\partial \bar{p}}{\partial x_i} + \nu \frac{\partial^2 \bar{u}_i}{\partial x_j \partial x_j} + f_{b_{i}} - \frac{\partial \tau_{ij} }{\partial x_j} \\
\frac{\partial \bar{u}_i}{ \partial x_i}  &=& 0, 
\end{eqnarray}

where \space $\bar{     }$ \space represents a low-pass spatially filtered quantity, $u_i$ are the three components of velocity, $p$ is pressure, $\nu$ is kinematic viscosity, and $\rho$ is density.  The sub-grid scale stresses are calculated with a constant Smagorinsky model, where

\begin{equation}
\label{e:sgs}
\frac{\partial \tau_{ij} }{\partial x_j} = -2 C_s^2 \Delta^2 |\bar{S}| \bar{S}_{ij}
\end{equation}

and the filtered rate of strain is

\begin{equation}
\label{e:strain}
\bar{S}_{ij} = \frac{1}{2}\left (\frac{\partial \bar{u}_i}{\partial x_j} + \frac{\partial \bar{u}_j}{\partial x_i}\right ).
\end{equation}

For all simulations the Smagorinsky constant is $C_s=0.1$. Rigid body motion is added by prescribing the appropriate body forces, 

\begin{eqnarray}
f_{bx}&=& -u_2 \dot{\theta} - x_1 \dot{\theta}^2 - x_2 \ddot{\theta}
\\
f_{by}&=& \ddot{h} + u_1 \dot{\theta} - x_2 \dot{\theta}^2 + x_1 \ddot{\theta}
\label{e:bodyforces}
\end{eqnarray}

\noindent
to account for the accelerations in the non-inertial frame of reference, where $\theta(t)$ is the instantaneous angle of inclination of the foil with respect to the positive $x_1$-axis, and $h(t)$ is the instantaneous vertical position of the foil during the heave cycle.

A second-order finite-volume solver is used to evaluate the flow around the oscillating hydrofoil. The solver is built from the {\it OpenFOAM} libraries, an open-source package for numerical algorithms and solvers. The libraries provide a framework of fully parallelized finite-volume based operators, but allow for user input for modification of boundary conditions, schemes, modeling parameters, and discretization techniques. A standard pressure-implicit split-operators (PISO) Navier-Stokes solver was modified to include the appropriate body force terms and a second-order backward time-stepping algorithm is utilized to solve for pressure on a collocated grid, without any relaxation parameters. 

\subsection{Mesh details}

\begin{figure}
\centering
\includegraphics[width=0.5\textwidth]{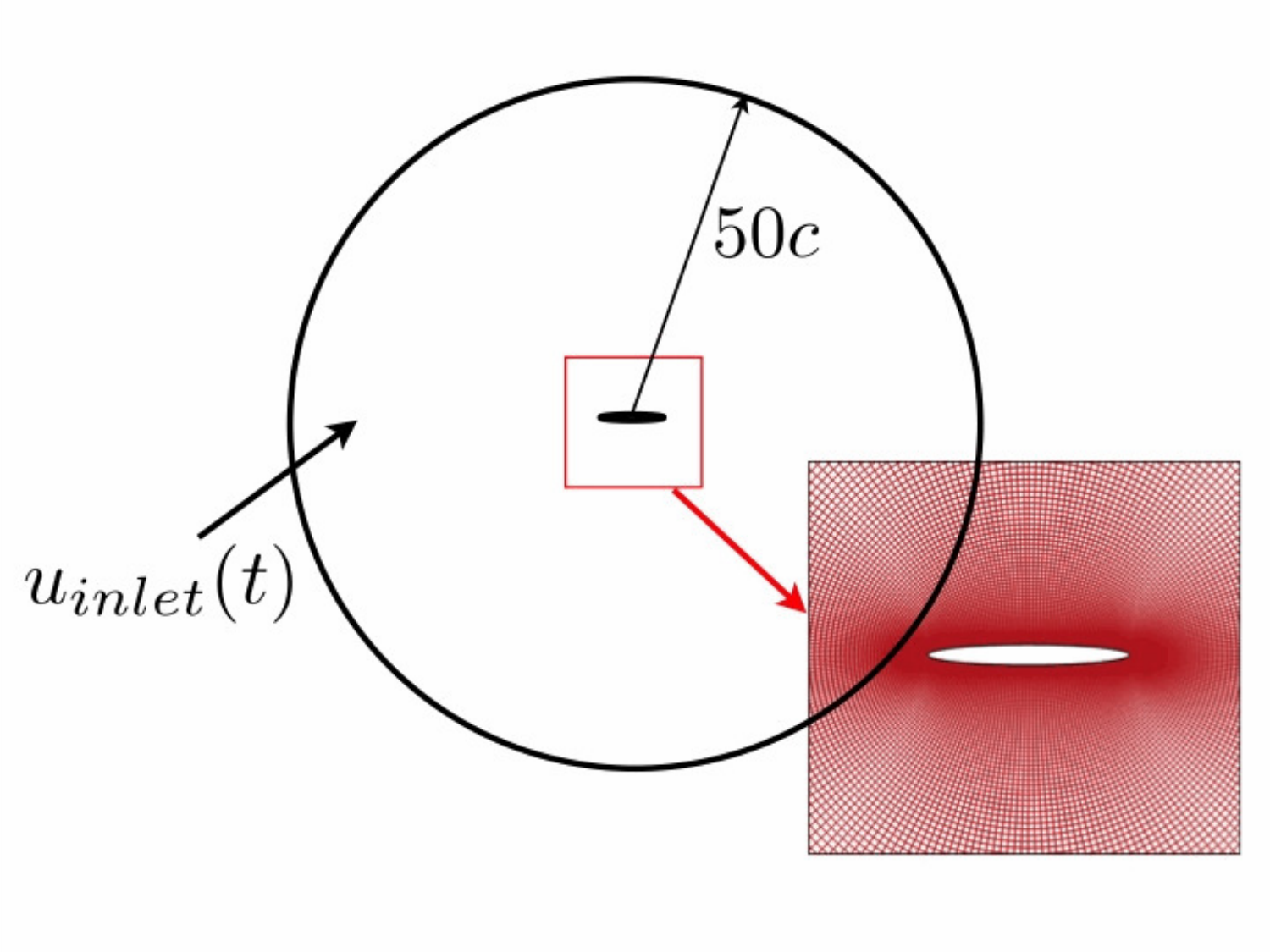}
\caption[Mesh Diagram]{Computational domain with local mesh surrounding ellipse, with points highly clustered around the hydrofoil for a well-resolved boundary layer.}
\label{mesh}
\end{figure}

For both the DNS and LES meshes, a conformal mapping routine is utilized to create an orthogonal mesh surrounding an ellipse of aspect ratio 10, with a radial boundary of 50 chord lengths ($50c$) in all directions (Fig. \ref{mesh}). The two-dimensional DNS mesh is comprised of 240 points equally spaced along the body, and 225 points in the radial direction using a tangent stretching function to cluster more points in the vicinity of the body.  A full mesh resolution study was performed to determine the minimal number of points in which the unsteady solution became independent of mesh resolution.

The final three-dimensional LES mesh is created similarly, but extruded spanwise ($z$ coordinate direction) for a domain width of $L_z=0.2c$. To ensure proper boundary layer resolution, the LES mesh has carefully clustered points within the boundary layer such that the local resolution at the wall is $\Delta y^+ <1$, where $y^+$ denotes the wall-normal coordinate system non-dimensionalized by the local friction velocity in the boundary layer for a stationary foil at the same Reynolds number. The spanwise direction is divided equally into 48 cells (49 nodes) over the domain of $L_z/c=0.2$ giving a value of $\Delta z/c=4.17 \times 10^{-3}$ (mesh 4 in Table \ref{tab:table3}).  

The boundary conditions are comprised of a no-slip condition on the foil and a time-varying inlet condition that adjusts the flow according to the local angle of attack of the foil throughout its pitch and heave cycle on the outer radius. A buffer region approximately 10 chord lengths in size is implemented at the outer boundary to remove superfluous numerical oscillations along the outer boundary. The flow is periodic in the spanwise direction.

A mesh independence study was performed on five meshes with varying resolution in the spanwise, radial and angular directions. Table \ref{tab:table3} shows the resolution, computed efficiency $\eta$ and wall-resolution, $\Delta y^{+}$.  The mesh points in the radial direction are distributed differently in each refinement, and thus a lower number of points can correspond to an additional increase in the near-wall resolution, as seen from the difference between mesh 1 and mesh 3.  The points are uniformly distributed in the spanwise direction.  Additionally, the Smagorinsky constant, $C_{s}$, was changed from $0.1$ to $0.15$ in order to assess the sensitivity of the turbulence model for mesh 3. The case analyzed has the kinematics $fc/U_{\infty}=0.1$,  $h_o/c=1.5$ and $\theta_o=65^{\circ}$.

\begin{table}[h]
\centering
\caption{Mesh details showing the total number of nodes (N), number of nodes in radial ($N_{r}$), angular ($N_{\theta}$), spanwise ($N_{z}$) directions, efficiency $\eta$ and near wall resolution $\Delta y^{+}$.}
\label{tab:table3}
\setlength{\tabcolsep}{12pt}
\begin{tabular}{lcccccc}
\hline \hline
Mesh             & $\eta$ & $N$ & $N_{r}$ & $N_{\theta}$ & $N_{z}$ & ${\Delta y^{+}}$ \\ \hline
Mesh 1          & 0.20              		& $1.78 \times 10^6$ & 225              & 240                 & 33                & 2.10                    \\
Mesh 2          & 0.23              		& $3.51 \times 10^6$ & 225              & 240                 & 65                & 2.69                    \\
Mesh 3           & 0.23                		& $1.19 \times 10^6$ & 150              & 240                 & 33                & 0.76                    \\
Mesh 4          & 0.22               		& $1.76 \times 10^6$ & 150             & 240                  & 49                & 0.79                    \\
Mesh 5           & 0.23              		& $2.34 \times 10^6$ & 150             & 240                  & 65                & 0.82                    \\ \hline \hline
\end{tabular}
\end{table}

\begin{figure}[h]
\centering
\includegraphics[width=0.58\textwidth]{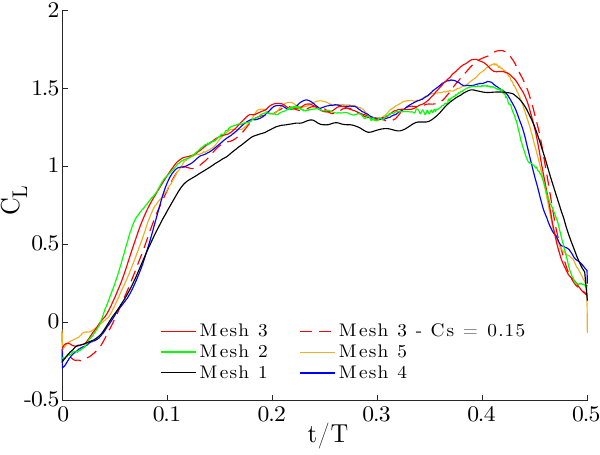}
\caption{Phase-averaged and span-averaged lift coefficient, $C_{L}$, during upstroke for various mesh resolutions for kinematics $fc/U_{\infty}=0.1$,  $h_o/c=1.5$, $\theta_o=65^{\circ}$.}
\label{fig:resolution}
\end{figure}


\begin{figure}[H]
\setlength{\fboxsep}{0pt}%
\setlength{\fboxrule}{0.1pt}%
\centering
    \begin{subfigure}{0.5\textwidth}
      	\makebox[\textwidth][c]{\includegraphics[width=1\linewidth]{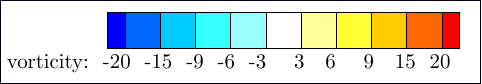}}
    \end{subfigure}
    
	\begin{subfigure}[b]{0.325\textwidth}
      	\fbox{\includegraphics[width=1\linewidth]{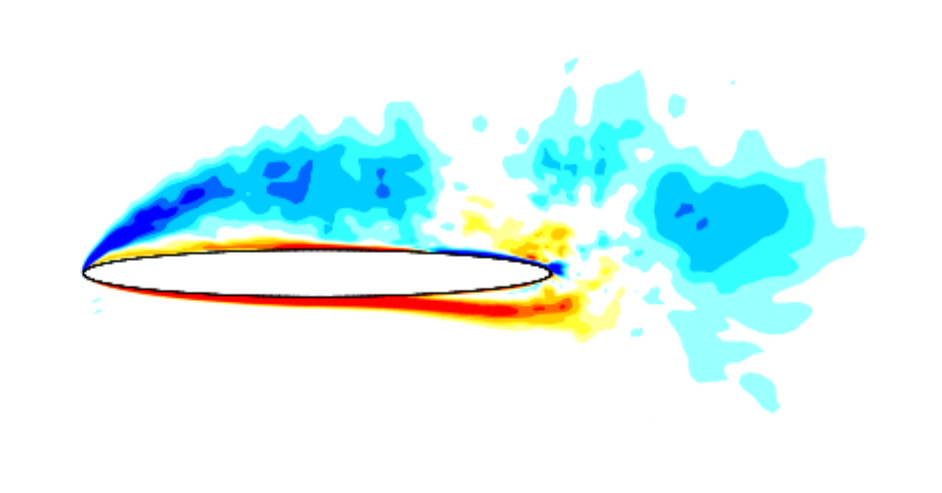}}
      	\caption{Mesh 1}
    \end{subfigure}
    \begin{subfigure}[b]{0.325\textwidth}
		\fbox{\includegraphics[width=1\linewidth]{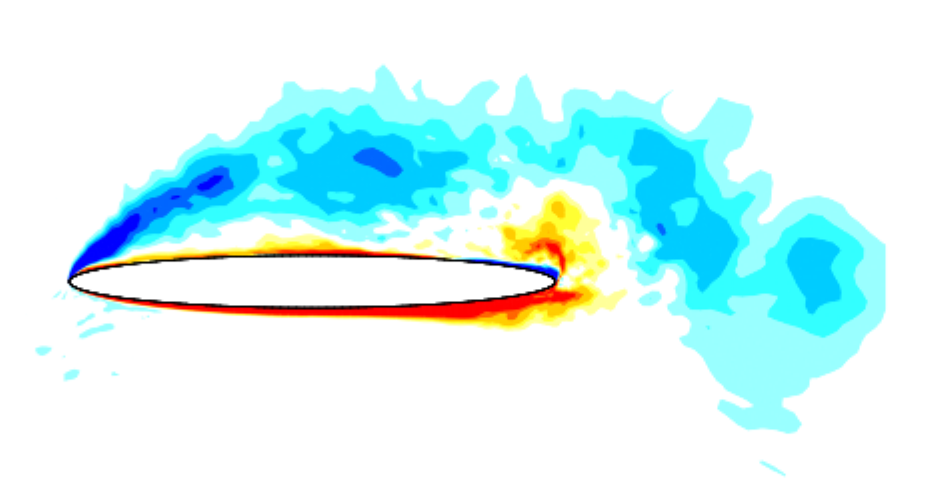}}
		\caption{Mesh 2}
	\end{subfigure}
	\begin{subfigure}[b]{0.325\textwidth}
		\fbox{\includegraphics[width=1\linewidth]{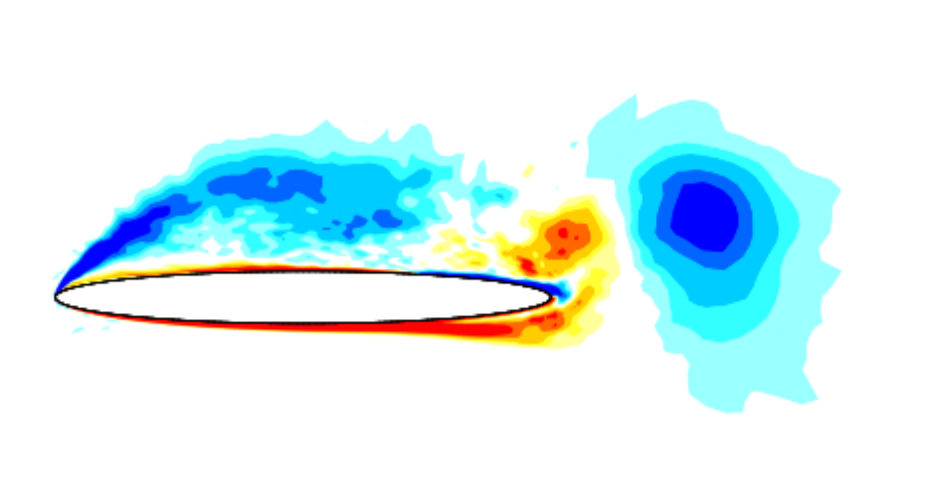}}
		\caption{Mesh 3 - Cs = 0.15}
	\end{subfigure}
	
	\begin{subfigure}[b]{0.325\textwidth}
		\fbox{\includegraphics[width=1\linewidth]{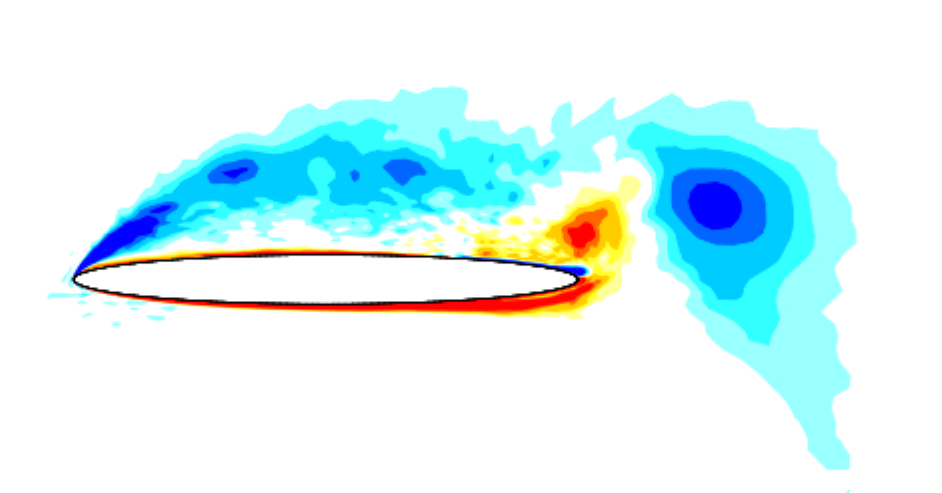}}
		\caption{Mesh 3}
	\end{subfigure}
	\begin{subfigure}[b]{0.325\textwidth}
		\fbox{\includegraphics[width=1\linewidth]{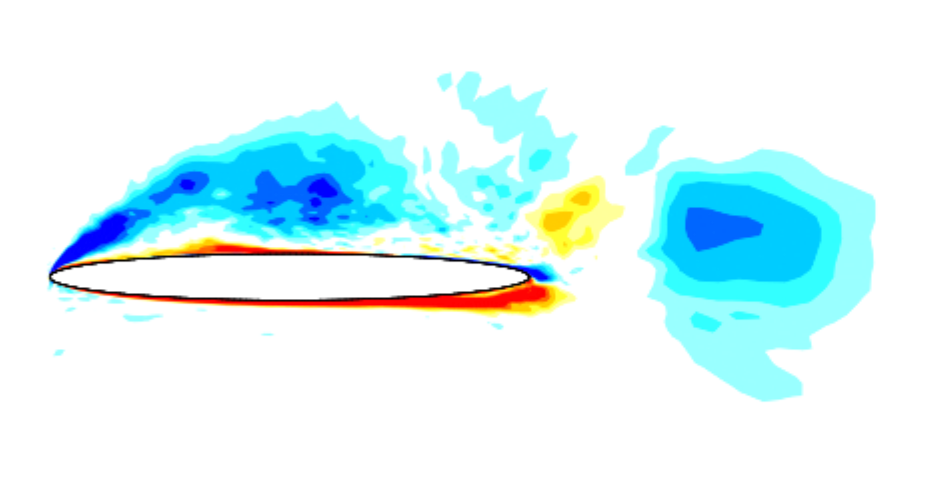}}
		\caption{Mesh 4}
	\end{subfigure}
	\begin{subfigure}[b]{0.325\textwidth}
		\fbox{\includegraphics[width=1\linewidth]{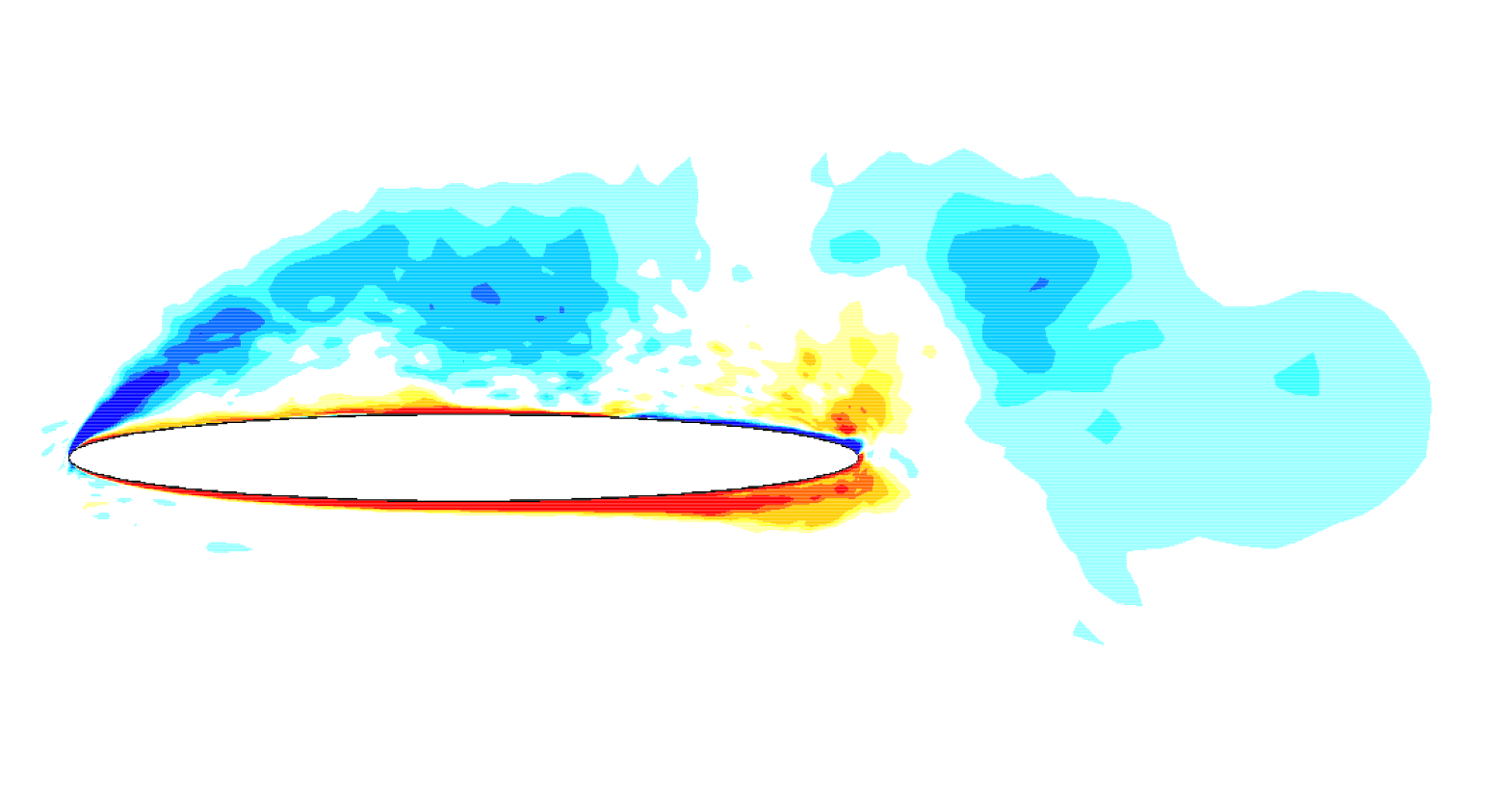}}
		\caption{Mesh 5}
	\end{subfigure}
\caption{Phase-averaged and span-averaged vorticity at top of upstroke for kinematics $fc/U_{\infty}=0.1$,  $h_o/c=1.5$ and $\theta_o=65^{\circ}$.}
\label{fig:meshing}
\end{figure}

The variation in terms of efficiency is not significant for meshes 2-5, however a significant difference in efficiency is noted with mesh 1, which is likely due to under-resolution in the boundary layer. To further examine the effect of the mesh, the span-averaged and phase-averaged lift coefficient is plotted in Fig. \ref{fig:resolution}, and the span-averaged and phase-averaged vorticity fields are shown in Fig. \ref{fig:meshing} for the cases displayed in Table \ref{tab:table3}. From this information it was deduced that mesh 3 is under-resolved in the spanwise direction due to the strength and location of the counter-clockwise rotating vortex at the trailing edge. Due to the high computational expense of mesh 5, and its similarity with mesh 4 in terms of the flow-field, efficiency, and lift forces, mesh 4 was selected to be the production mesh for the LES.

\subsection{Simulation parameters}

The kinematic motion of the hydrofoil is prescribed through the body force term in Eq. \ref{e:bodyforces} and is described below in lab-fixed coordinates as

\begin{equation}
h(t)=h_{o}\cos(2\pi f t) 
\label{eq:heave}
\end{equation}

and 

\begin{equation}
\theta(t)=\theta_{o}\cos(2\pi f t + \pi/2)
\label{eq:pitch}
\end{equation}

where $h_{o}$ is the maximum heave amplitude, $\theta_o$ is the maximum pitch amplitude and $f$ is the non-dimensional frequency, normalized by chord length and freestream velocity. The pitch and heave strokes always use the same frequency and are always separated by a phase difference of $\pi/2$, which is found to yield the optimal power \citep{zhu2011}.  Modifying pitch and heave velocities simultaneously yields a time-varying relative angle of attack of the foil with respect to the freestream flow, which can be computed with

\begin{equation}
\alpha_{rel}(t) = \tan^{-1} (-\dot{h}(t)/U_{\infty}) + \theta(t).
\label{eq:alpharel}
\end{equation} 

Due to the periodic nature of the kinematics, time is represented by percent of the total cycle, $t/T$, where $T$ is the period of oscillation. The cycle starts at $t/T=0$ when the foil is at the bottom of the heave stroke, and orientated with a zero relative angle of attack. The representative relative angle of attack is evaluated at $t/T = 0.25$, when the heave velocity is at a maximum, and is given as

\begin{equation}
\alpha_{T/4} =\alpha_{rel}(t = 0.25T).
\label{eq:at4}
\end{equation}

In order to evaluate the performance of different kinematic conditions, the efficiency is defined as

\begin{equation}
\eta = \frac{\bar{P}}{\frac{1}{2}\rho U_{\infty}^3 Y_p}
\label{eq:eta}
\end{equation}

\noindent
which is the ratio of the average power extracted throughout a single stroke, $\bar{P}$, compared to the power available in the freestream velocity throughout the swept area $Y_p$.  Power is defined as

\begin{equation}
P(t)=F_y \dot{h} + M_z \dot{\theta}
\label{eq:power}
\end{equation}

\noindent
where $F_y$ and $M_z$ are the vertical force and spanwise moment on the foil respectively. Thus the power is comprised of a translational contribution, $F_y \dot{h}$, and an angular contribution, $M_z \dot{\theta}$. 

The calculation of efficiency includes the total available power from the fluid and the total swept area of the device, $Y_p$, which increases with $h_o/c$. The swept area $Y_p$ is often greater than twice the heave amplitude, since it takes into account the largest area swept (see Fig. \ref{f:stroke}). Another parameter of interest to the renewable energy community is the maximum power of the device regardless of the kinematic stroke. Here, this parameter is defined by the power coefficient, $C_p$, 

\begin{equation}
C_p = \frac{\bar{P}}{\frac{1}{2}\rho U_{\infty}^3 Sc}
\label{eq:cp}
\end{equation}

where the denominator is fixed to the chord and span of the foil, $S$, and does not change with varying kinematics. To remove small cycle-to-cycle variations as best as possible, the efficiency, power coefficients, forces, and flow fields are all phase-averaged through the last 6 half-cycles of simulation, and the LES data is also span-averaged.

The computations included in this paper will include a parameter sweep through various heave amplitudes ($h_o/c=0.5-2$) and pitch amplitudes ($\theta_o=60^{\circ}-95^{\circ}$)  at two non-dimensional frequencies of $fc/U_{\infty}=0.1$ and $fc/U_{\infty}=0.15$.  Based on previous data [\citealp{Xiao2014,kindum2008, Young2014}], the global maximum efficiency, $\eta$, and maximum power coefficient, $C_p$, are believed to exist within the parameter space tested above.

\section{Results and discussion}

\subsection{Leading edge vortex and effect of kinematics}

\begin{figure}[H]
\centering
	\begin{subfigure}{0.49\textwidth}
	\centering
      	\includegraphics[width=1\linewidth]{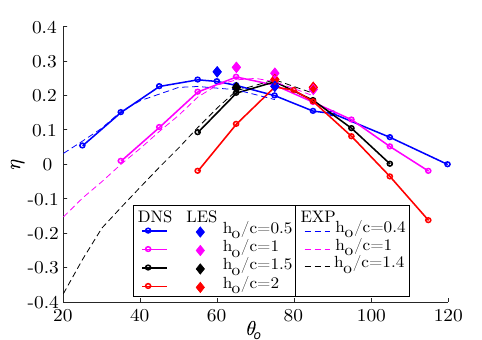}
		\caption{$fc/U_{\infty}=0.1$}
	\end{subfigure}
	\begin{subfigure}{0.49\textwidth}
	\centering
		\includegraphics[width=1\linewidth]{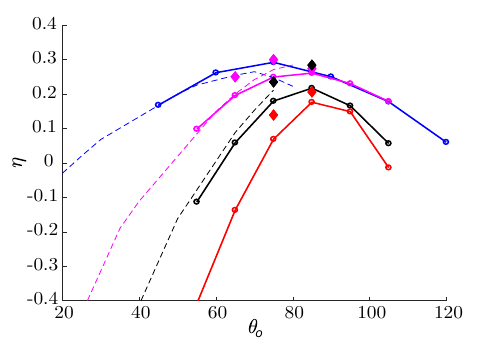}
		\caption{$fc/U_{\infty}=0.15$}
	\end{subfigure}
\caption{The efficiency, $\eta$, as a function of pitch amplitude, $\theta_o$, for a series of heave amplitudes $h_o/c$. DNS (---) at $Re=1000$ is directly compared to experimental flume data (- - -) at $Re=50,000$ and LES data (diamonds) at $Re=50,000$. }
\label{fig:etaexp}
\end{figure}

Different kinematics from the simulations are directly compared with available flume data \citep{Kim2017} in Fig. \ref{fig:etaexp}, in which the efficiency for reduced frequencies $fc/U_{\infty}=0.1$ and $0.15$ are displayed as a function of pitch and heave amplitude.

The LES data in Fig. \ref{fig:etaexp} closely matches the experimental flume conditions, which has an elliptical foil shape pitched about the center chord and a freestream Reynolds number of $Re = 50,000$ \citep{Kim2017}. The DNS maintains the same flow conditions performed at a lower Reynolds number of $Re = 1000$.

The maximum efficiency in the heave range of $h_o/c=0.5-2$ is 28.1\% for $fc/U_{\infty}=0.1$, and $30.0\%$ for $fc/U_{\infty}=0.15$, both at the higher Reynolds number. Both peaks occur at $h_o/c=1$, with pitch amplitudes of $65^{\circ}$ and $75^{\circ}$, respectively. For both frequencies tested there is a sharp drop-off in efficiency with higher and lower pitch amplitudes (holding heave amplitude constant), and the optimal pitch amplitude shifts from approximately $60^{\circ}$ with the lowest heave amplitude of $h_o/c=0.5$, to approximately $80^{\circ}$ with the highest heave amplitude of $h_o/c=2$. These trends and peak efficiency values are consistent with the literature [\citealp{kindum2008,kinsey2012,Kim2017,Simeski2017,SuThesis,MillerThesis,zhu2011,campobasso2013,Ashraf2011,Xiao2012}].

\begin{figure*}[!htbp]
        \centering
        \setlength\tabcolsep{1.5pt}
        \begin{tabular}{cccccc}
             & $h_o/c=0.5$ & $h_o/c=1.0$ & $h_o/c=1.5$ & $h_o/c=2.0$ \\
  	    \rotatebox{90}{t/T=0.12} & 
            \includegraphics[trim=3.3cm 4cm 2.64cm 4.24cm, clip=true, width=0.22\textwidth]{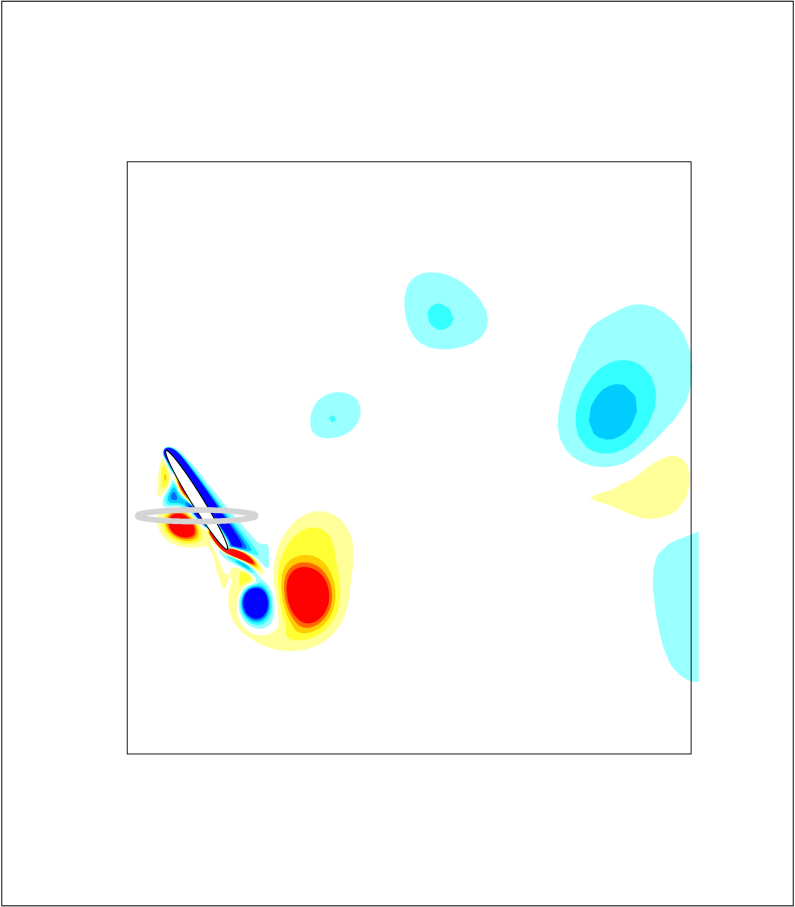} & 
            \includegraphics[trim=3.3cm 4cm 2.64cm 4.24cm, clip=true, width=0.22\textwidth]{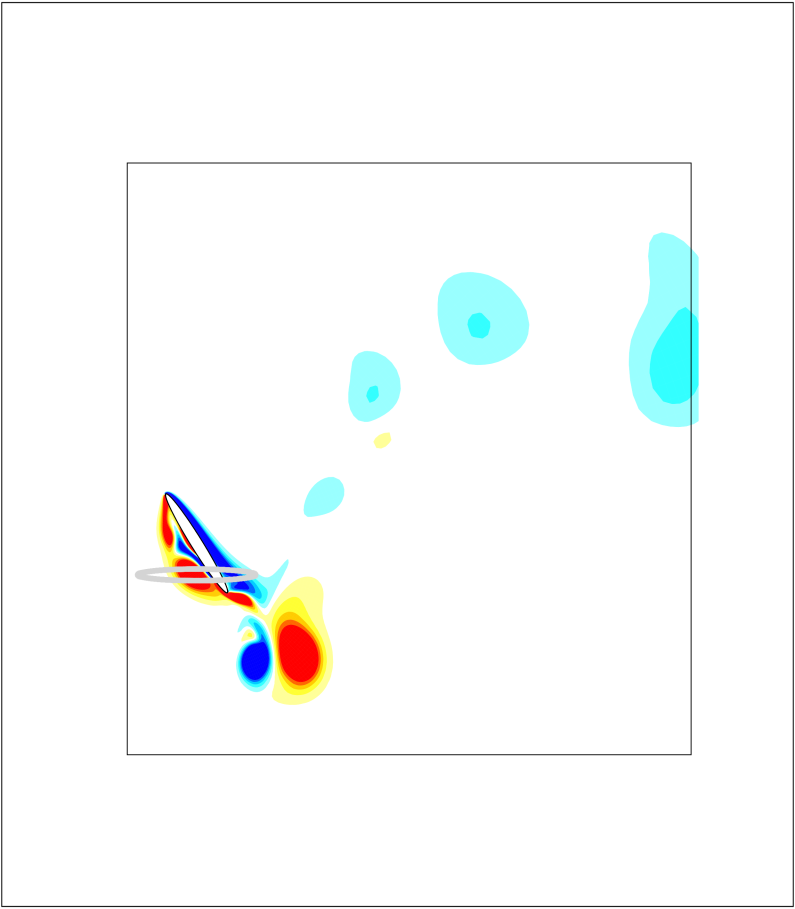} & 
            \includegraphics[trim=3.3cm 4cm 2.64cm 4.2cm, clip=true, width=0.22\textwidth]{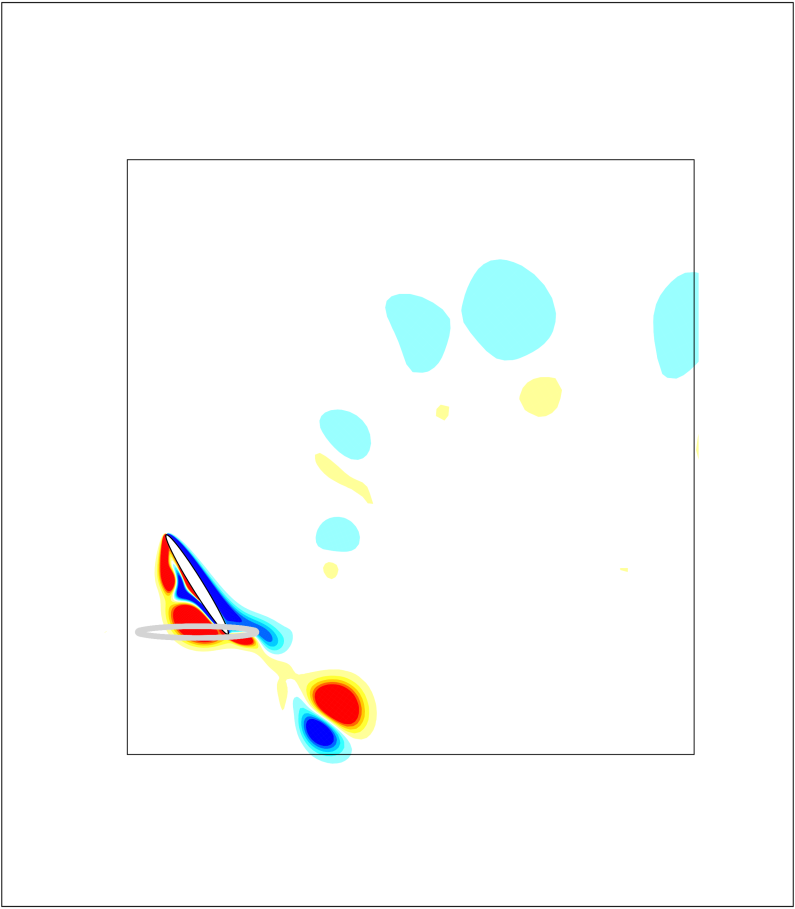} &
            \includegraphics[trim=3.3cm 4cm 2.64cm 4.24cm, clip=true, width=0.22\textwidth]{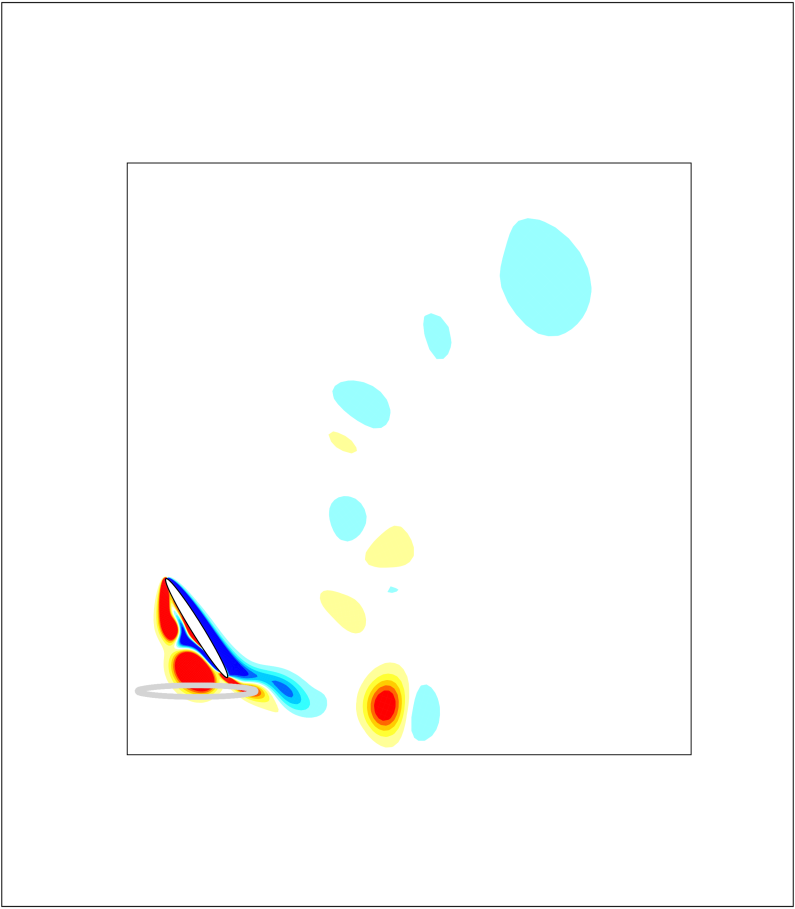} \\
            \rotatebox{90}{t/T=0.25} & 
            \includegraphics[trim=3.3cm 4cm 2.64cm 4.24cm, clip=true, width=0.22\textwidth]{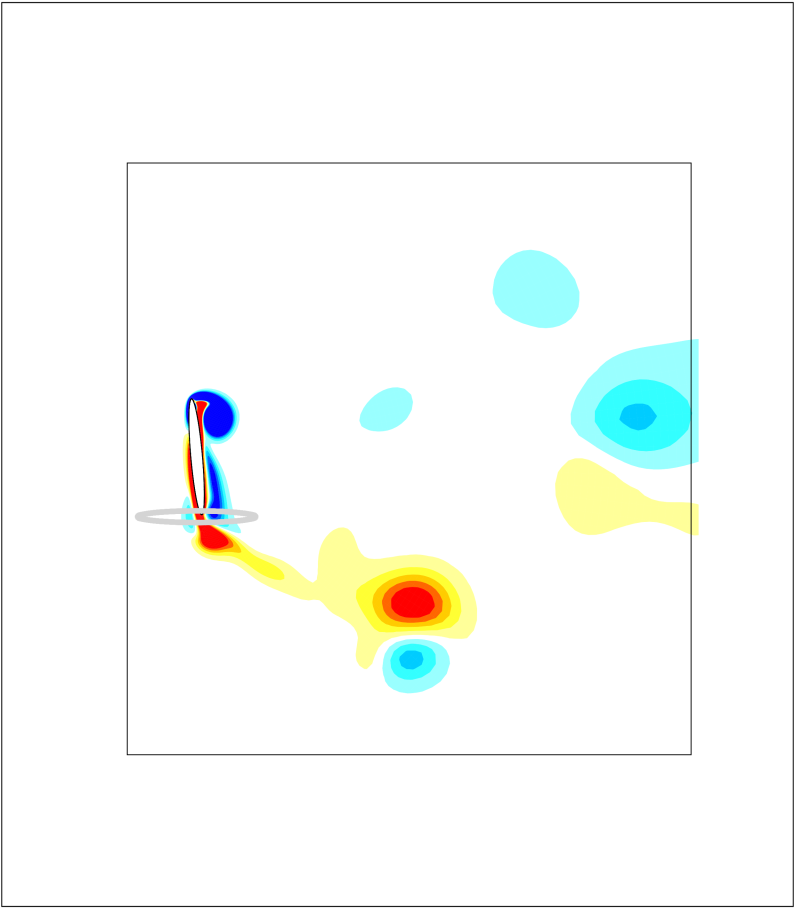}  & 
            \includegraphics[trim=3.3cm 4cm 2.64cm 4.24cm, clip=true, width=0.22\textwidth]{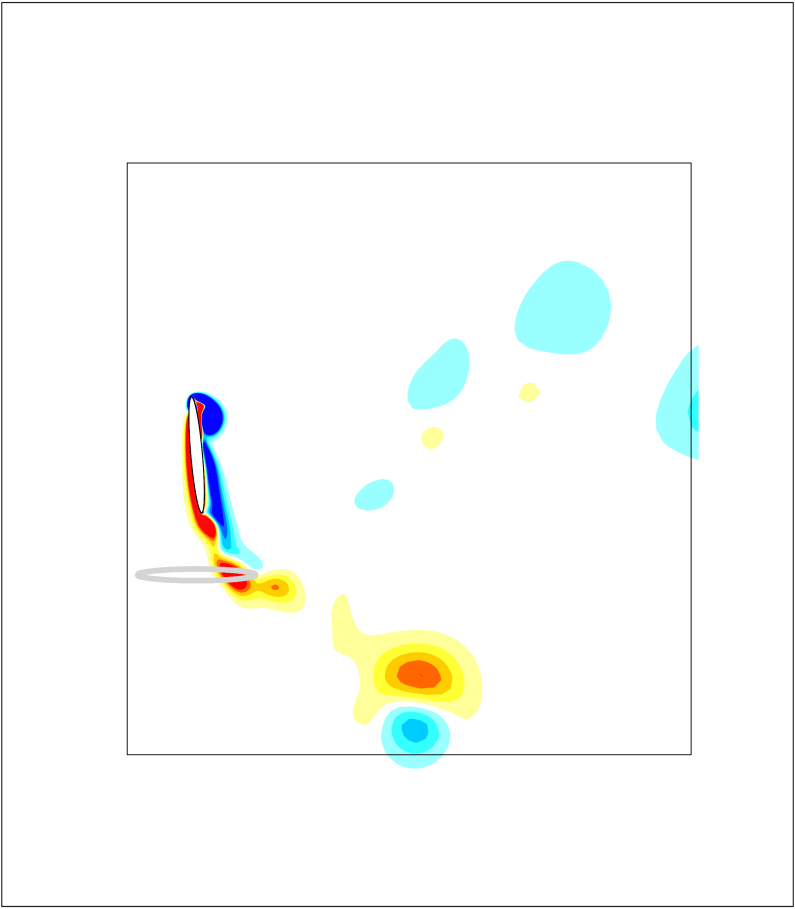} & 
            \includegraphics[trim=3.3cm 4cm 2.64cm 4.2cm, clip=true, width=0.22\textwidth]{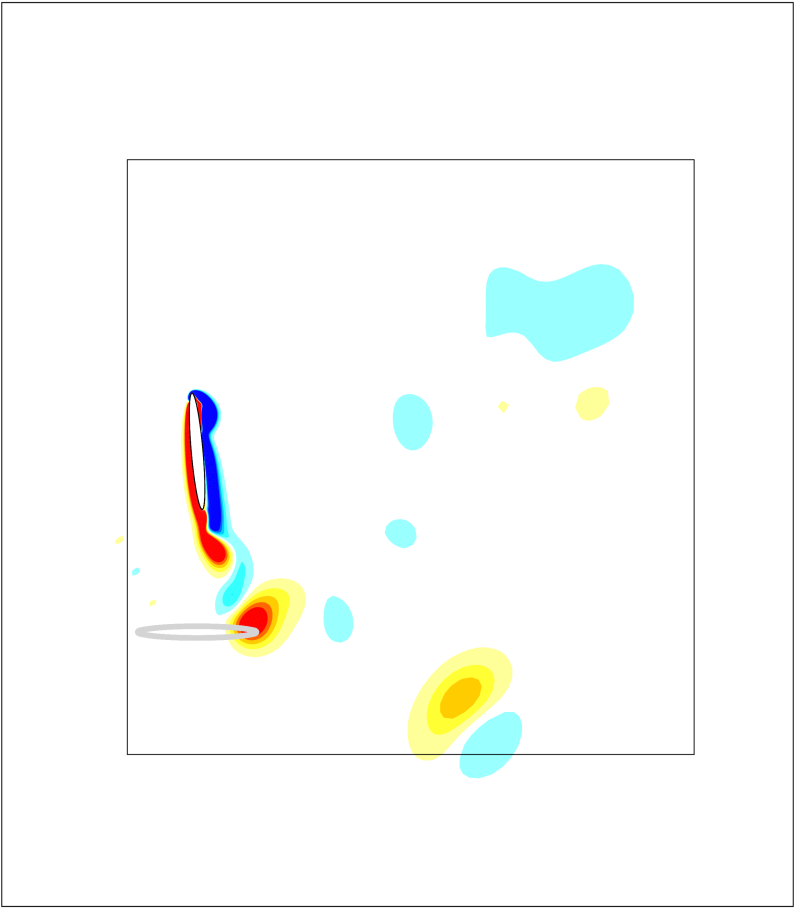} &
            \includegraphics[trim=3.3cm 4cm 2.64cm 4.24cm, clip=true, width=0.22\textwidth]{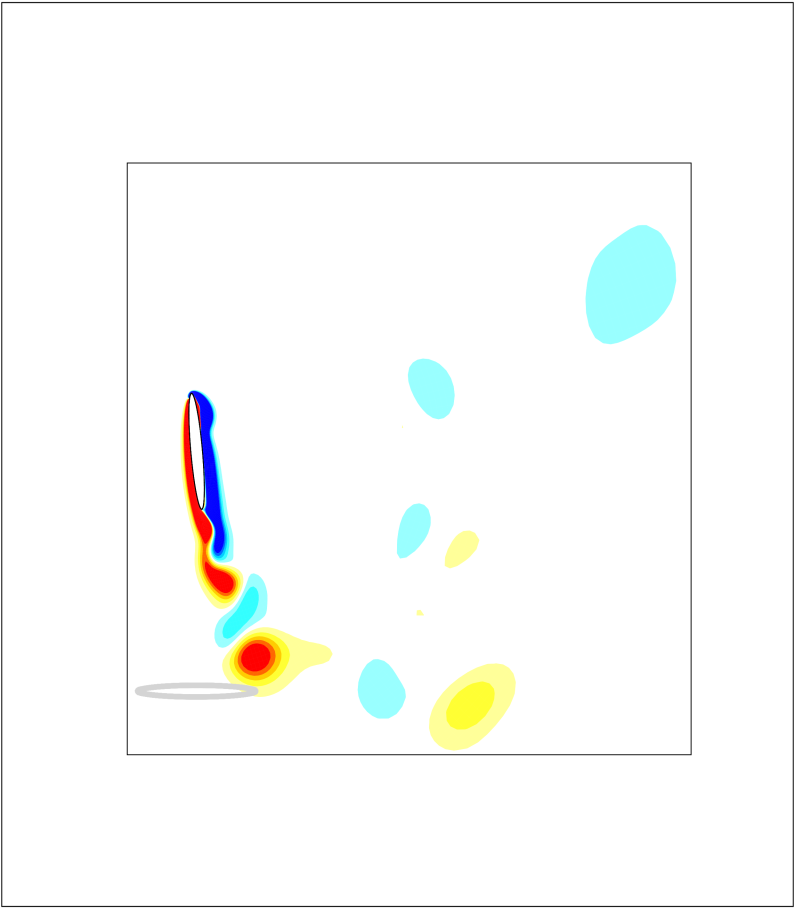} \\			
            \rotatebox{90}{t/T=0.38} &             
            \includegraphics[trim=3.3cm 4cm 2.64cm 4.24cm, clip=true, width=0.22\textwidth]{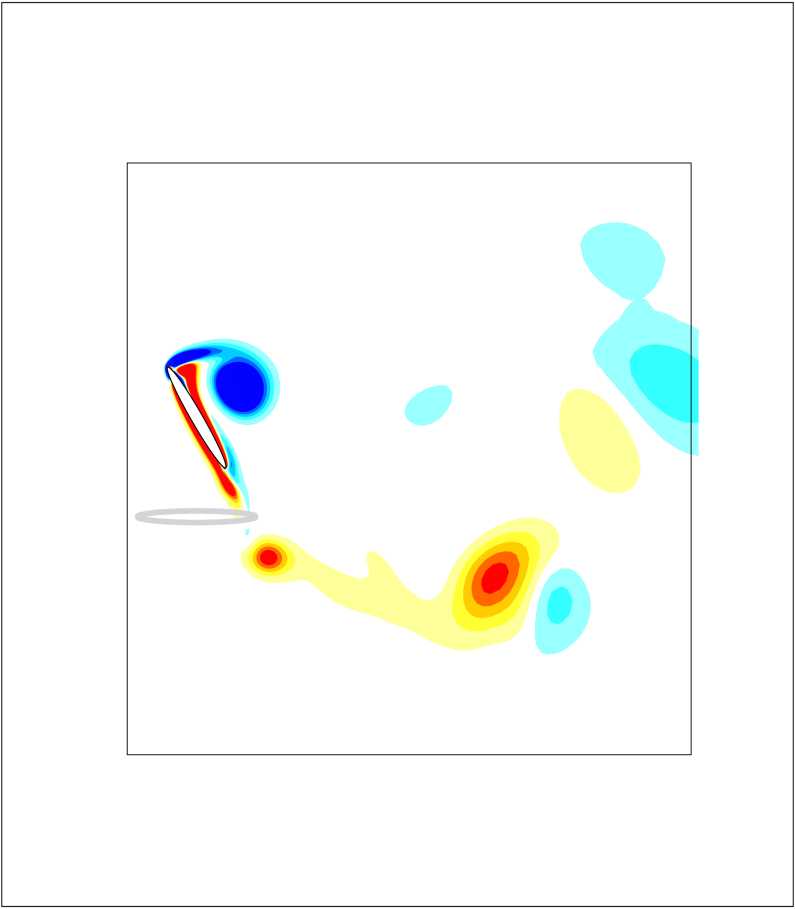}  & 
            \includegraphics[trim=3.3cm 4cm 2.64cm 4.24cm, clip=true, width=0.22\textwidth]{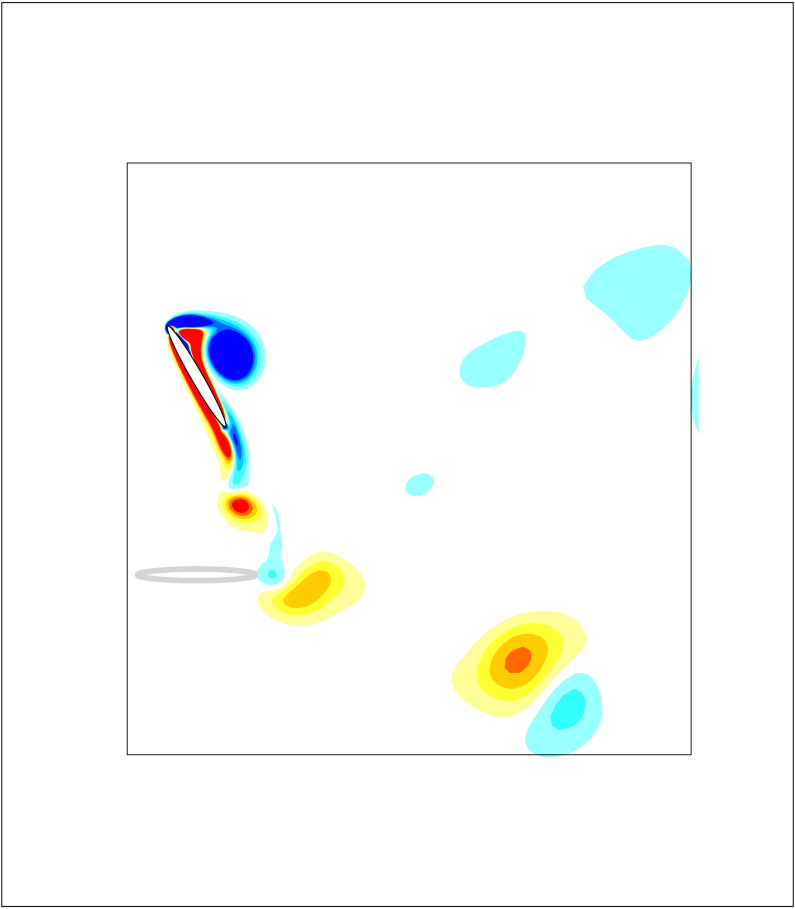} & 
            \includegraphics[trim=3.3cm 4cm 2.64cm 4.2cm, clip=true, width=0.22\textwidth]{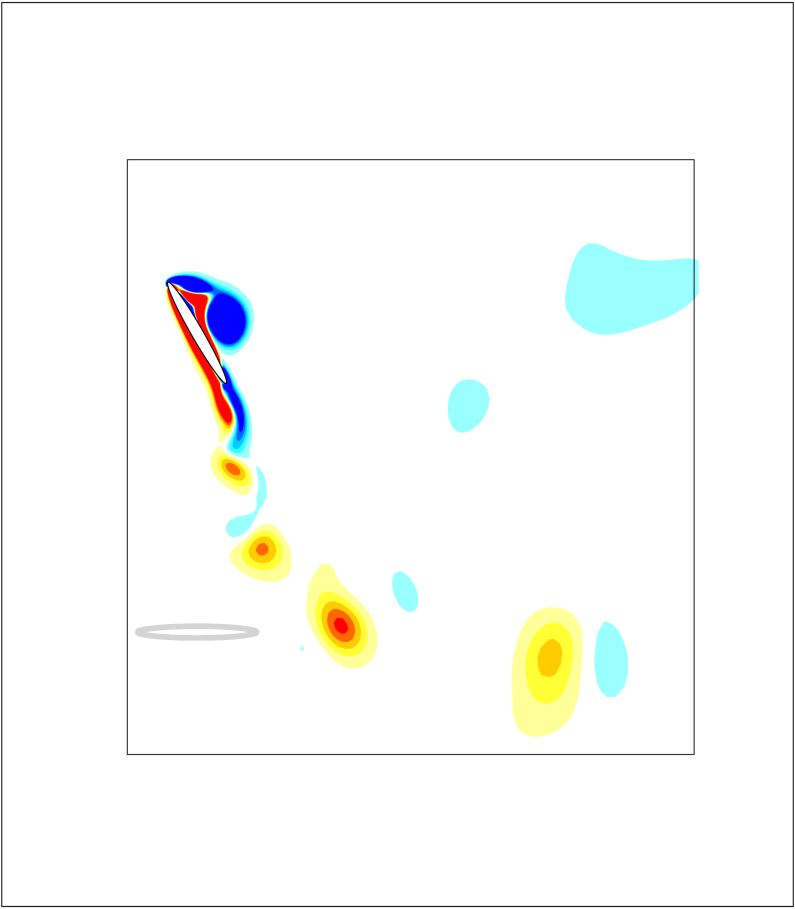} &
            \includegraphics[trim=3.3cm 4cm 2.64cm 4.24cm, clip=true, width=0.22\textwidth]{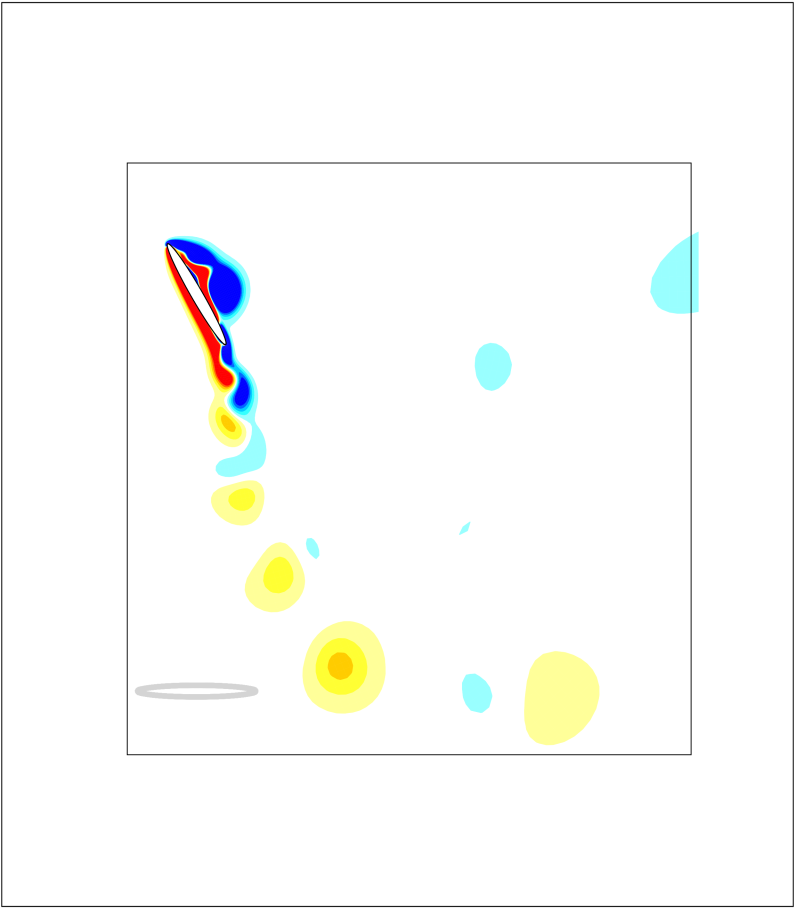} \\
            \rotatebox{90}{t/T=0.5} &             
            \includegraphics[trim=3.3cm 4cm 2.64cm 4.24cm, clip=true, width=0.22\textwidth]{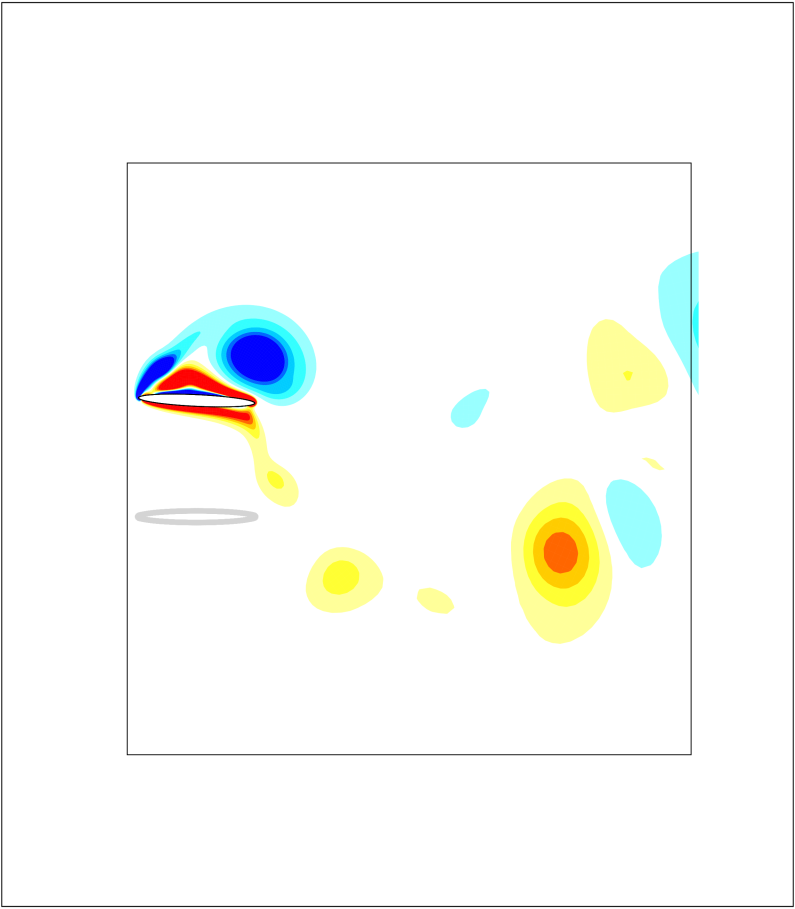}  & 
            \includegraphics[trim=3.3cm 4cm 2.64cm 4.24cm, clip=true, width=0.22\textwidth]{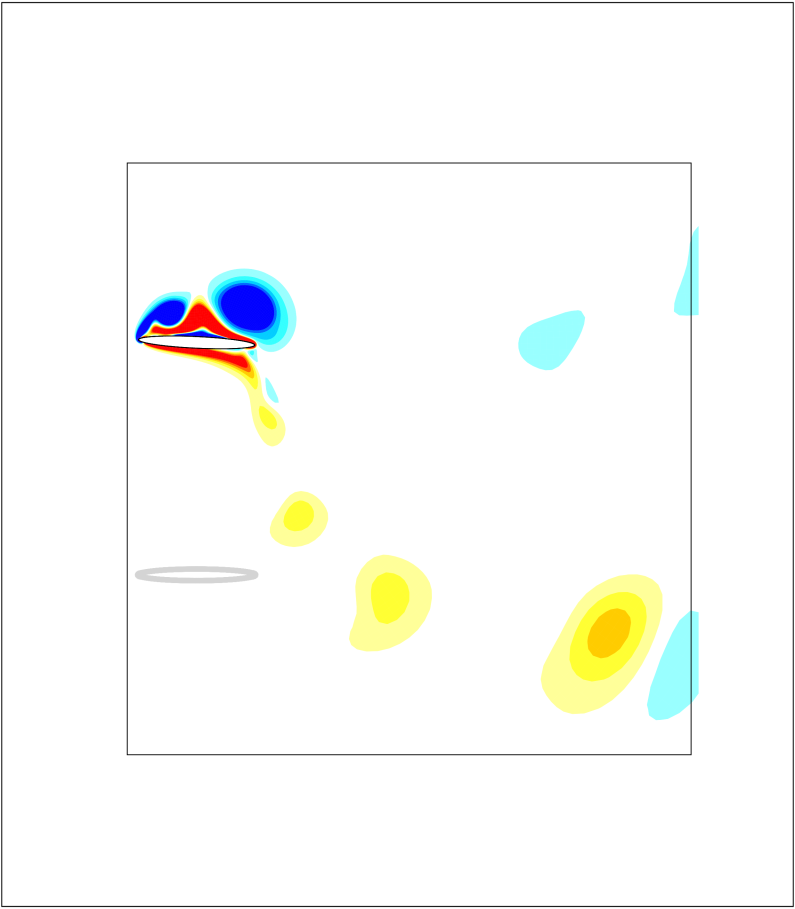} & 
            \includegraphics[trim=3.3cm 4cm 2.64cm 4.2cm, clip=true, width=0.22\textwidth]{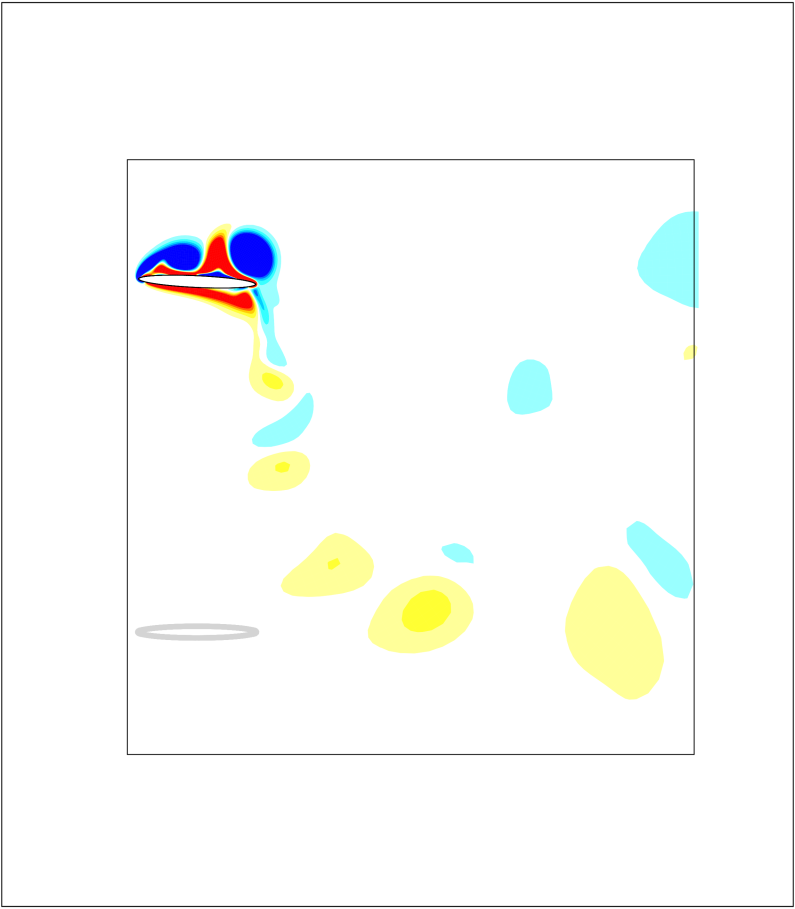} &
            \includegraphics[trim=3.3cm 4cm 2.64cm 4.24cm, clip=true, width=0.22\textwidth]{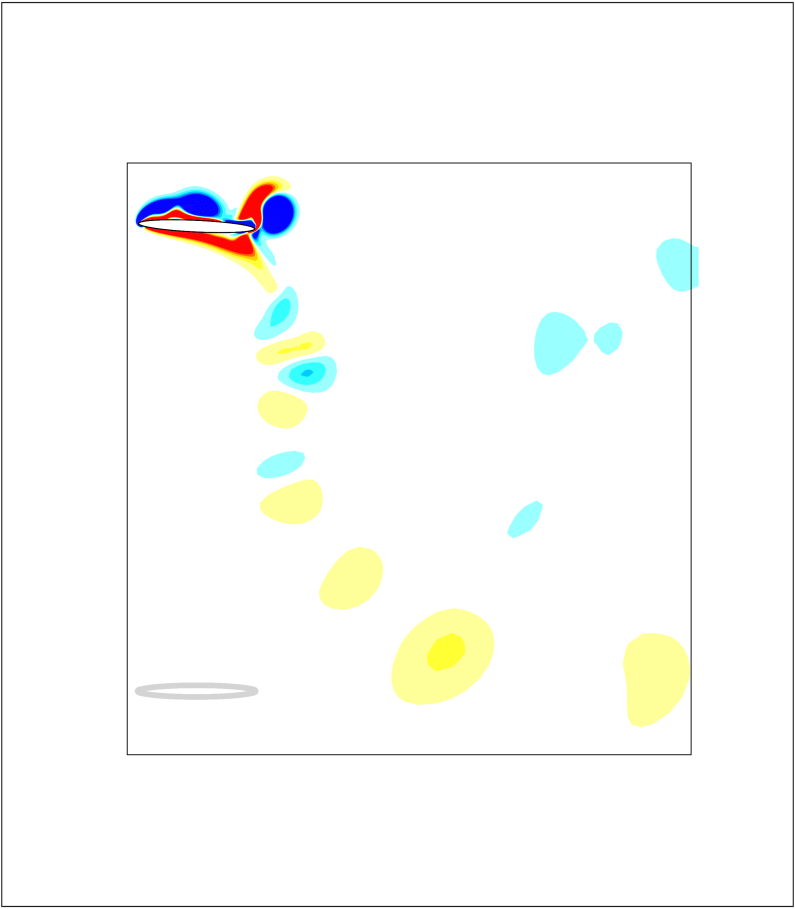} \\
        \end{tabular}
    \begin{subfigure}{0.5\textwidth}
      	\makebox[\textwidth][c]{\includegraphics[width=1\linewidth]{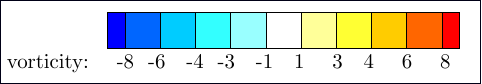}}
    \end{subfigure}
\caption{Progression of the LEV during upstroke for four heave amplitudes at $fc/U_{\infty}=0.15$ and $\theta_o=85^{\circ}$ for DNS at $Re=1000$.}
\label{fig:2dvortex}
\end{figure*}

Fig. \ref{fig:2dvortex} displays the vortex dynamics over an upstroke cycle across the heave amplitude range explored in this paper. The frequency and pitch amplitude are held constant at $fc/U_{\infty}=0.15$ and $\theta_o=85^{\circ}$ and $h_o/c$ ranges from $0.5$ to $2$. The gray foil is the bottom of the stroke which provides a reference for how far the foil must travel to complete the half-stroke translation. For each set of kinematics there is a prominent LEV that is formed at or before mid-upstroke ($t/T=0.25$). Within the high efficiency energy harvesting regime the formation time, position and size of the LEV is a non-linear function of all kinematic parameters; frequency, pitch amplitude, and heave amplitude.

\begin{figure}[!htbp]
\centering
\includegraphics[width=0.71\textwidth]{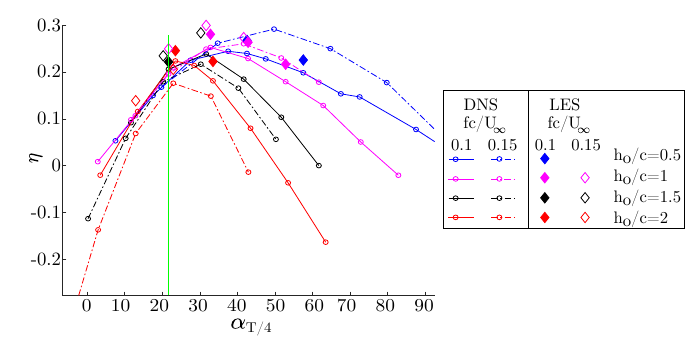}
\caption{Efficiency data from DNS and LES in Fig. \ref{fig:etaexp} rescaled as a function of the relative angle of attack, $\alpha_{T/4}$ (in degrees).}
\label{fig:alphat4}
\end{figure}

Due to the interdependence of all kinematic parameters, the data from Fig. \ref{fig:etaexp} is plotted as a function of relative angle of attack at mid-upstroke in Fig. \ref{fig:alphat4}. Across various frequencies and heave amplitudes, the efficiency data collapses for $\alpha_{T/4}<22^{\circ}$. In this regime the boundary layer is more or less attached to the foil with very little to no separation and no distinct LEV. The prominent LEV that contributes to the high efficiency modes begins when $\alpha_{T/4}>22^{\circ}$. At each heave amplitude the maximum efficiency is achieved within this range where $\alpha_{T/4}>22^{\circ}$.  Furthermore as the relative angle of attack increases beyond $22^{\circ}$, the efficiency becomes a strong function of heave amplitude, with the lowest heave amplitudes maintaining the highest overall efficiency with increasing $\alpha_{T/4}$, and the highest heave amplitudes dropping off sharply towards zero efficiency as $\alpha_{T/4}$ increases. Of the kinematics explored the maximum efficiency occurs within the range $35^{\circ}<\alpha_{T/4}<50^{\circ}$.

\subsection{Comparison with experimental data}

\begin{figure}[H]
\centering
      	\includegraphics[width=1\textwidth]{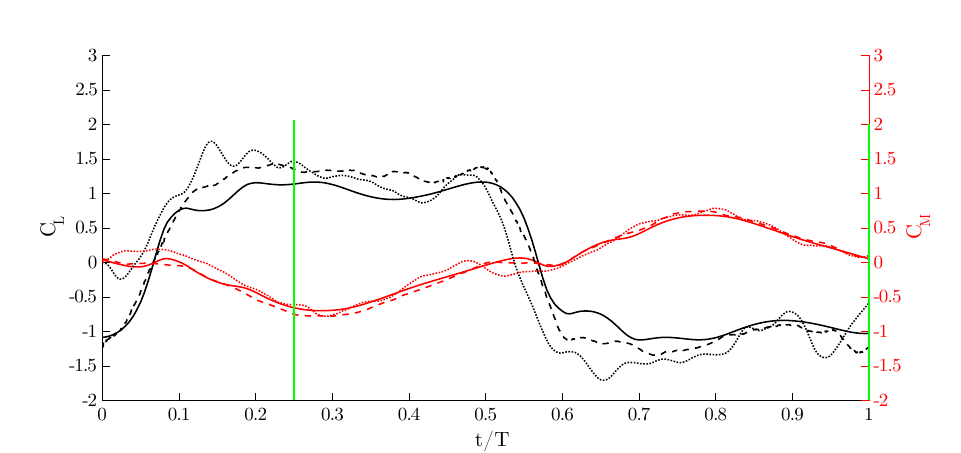}
		\caption{Time-dependent lift ($C_{L}$) and moment ($C_{M}$) coefficients for kinematics $fc/U_{\infty}=0.15$,  $h_o/c=1$, $\theta_o=65^{\circ}$. LES at $Re=50,000$ (- - -). DNS at $Re=1000$ (---). EXP at $Re=30,000$ (...). Cycle begins at bottom of stroke, upstroke from $t/T=0-0.5$ and downstroke from $t/T=0.5-1$. The vorticity fields represented by the green lines at mid-upstroke and end of downstroke can be seen in Fig. \ref{fig:vorticity}.}
		\label{fig:expcfdforcesf15}
\end{figure}

\begin{figure}[H]
\centering
    \begin{subfigure}{0.5\textwidth}
      	\makebox[\textwidth][c]{\includegraphics[width=1\linewidth]{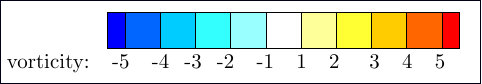}}
    \end{subfigure}
    
    \setlength\tabcolsep{1.5pt}
    \begin{tabular}{cccc}
    \rotatebox{90}{t/T=0.25} &
      	\includegraphics[width=0.28\linewidth]{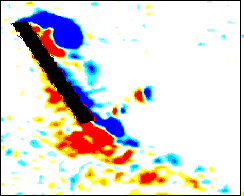} &
		\includegraphics[width=0.28\linewidth]{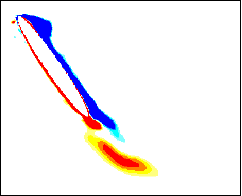} &
		\includegraphics[width=0.28\linewidth]{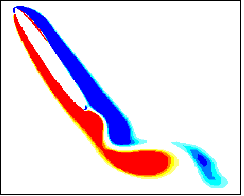} \\

	\rotatebox{90}{t/T=1} &	
		\includegraphics[width=0.28\linewidth]{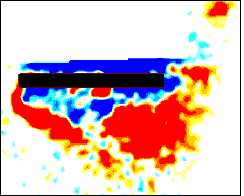} &
      	\includegraphics[width=0.28\linewidth]{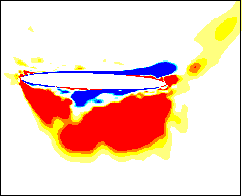} &
		\includegraphics[width=0.28\linewidth]{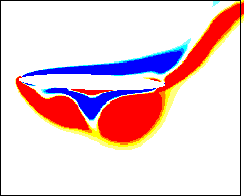} \\
		     & EXP; $Re=30,000$ & LES; $Re=50,000$ & DNS; $Re=1,000$
	\end{tabular}
\caption{Phase-averaged and span-averaged vorticity comparing experimental PIV (EXP), LES and DNS for kinematics $fc/U_{\infty}=0.15$, $h_o/c=1$, $\theta_o=65^{\circ}$. Experiments are performed with a flat-plate foil geometry whereas the simulations utilize an elliptical foil. Times shown are represented by the green lines in Fig. \ref{fig:expcfdforcesf15}.}
\label{fig:vorticity}
\end{figure}

Fig. \ref{fig:etaexp} demonstrates a strong agreement between experiments and simulations, and also no significant Reynolds number dependence within the range $Re = 1000-50,000$. The in-cycle forces and flow fields are also examined to better understand the similarities and differences between the LES, DNS, and experimental data. Figs. \ref{fig:expcfdforcesf15} and \ref{fig:vorticity} directly compare the kinematics of $fc/U_{\infty}=0.15$,  $h_o/c=1$ and $\theta_o=65^{\circ}$ in terms of phase-averaged forces, moments, and vorticity fields. In these comparisons the experiments are performed at $Re=30,000$ with a flat-plate geometry \citep{SuThesis}, and the conditions for the simulations are the same as Fig. \ref{fig:etaexp}.
Despite the slightly different experimental configuration, the experiments and simulations still show very good agreement in time-dependent lift ($C_L$) and moment ($C_M$) coefficients.

In Fig. \ref{fig:expcfdforcesf15}, the bottom of the downstroke is at $t/T=0$, and the upstroke occurs from $t/T=0-0.5$. The moment and lift coefficients switch sign at the top of the stroke or just after the stroke reversal, respectively, and repeat the behavior on the downstroke with reversed signs. The green lines at mid-upstroke ($t/T=0.25$) and the end of downstroke ($t/T=1$) have their respective vorticity fields presented in Fig. \ref{fig:vorticity} directly comparing the flow fields for experiments and simulations. 

The top row of Fig. \ref{fig:vorticity} shows the high-lift configuration at $t/T=0.25$, which consists of a clearly defined LEV in the LES and PIV vorticity fields. In the DNS the LEV is not yet formed but it does emerge at a slightly later time in the cycle. Due to this delay, the DNS has a lower lift coefficient in this portion of the cycle when compared with the higher Reynolds number LES and experimental results. This LEV persists along the suction surface of the foil until the stroke reversal, and it is present in all the high-efficiency kinematics explored in the paper. A strong positive lift force during upstroke, and a strong negative lift force during downstroke is responsible for most of the power generation, and results in power extracted from the translational, or linear motion of the foil, $C_{p,lin}$.  An opposite sign (counter-clockwise rotating) LEV is shown in the bottom row of Fig. \ref{fig:vorticity} just before it is shed at the bottom of the downstroke. The precise formation time and shedding cycle of the LEV will vary as a function of kinematics, as discussed in more detail in Section \ref{wakedynamics}. It is found that the energy harvesting efficiency of the foil is very strongly correlated to the LEV formation and shedding time which is governed almost exclusively by the kinematics of the foil and only a weak function of Reynolds number. Likewise, previous experiments have shown that the foil geometry has only a minor effect on the efficiency so long as the leading edge is sharp enough to produce an LEV \citep{Kim2017}.

\subsection{Contribution of angular power} \label{angpower}

In addition to supplementing the high-lift configuration, the LEV in the top row of Fig. \ref{fig:vorticity} creates a negative pitching moment. While the foil is increasing its pitch angle (positive angular velocity) the negative pitching moment contributes negative angular power extraction ($C_{p,ang}<0$). As pitch angle decreases and the angular velocity sign becomes negative (from $t/T=0.25-0.75$) the power due to angular motion is positive. Foil kinematics highly influence the LEV convection and time spent on the foil. At high frequency or high heave amplitude the LEV remains attached longer and has more influence on the vertical force and moment, both of which contribute to power. When the LEV convects past the pitching point at mid-chord, there is a strong negative pitching moment at $t/T=0.25-0.50$ in the opposite direction of the angular velocity, thus creating a negative contribution to power. Thus, as noted in previous experimental studies [\citealp{Kim2017,Su2019}], the exact location and timing of the LEV has a strong influence on the angular power. However, due to the relatively small values of torque compared to lift, the angular power contribution is more often smaller than the power extracted from the translational motion.

In order to further investigate the role of Reynolds number, and the effect of angular power contributions in high-efficiency configurations, Table \ref{tab:table2} directly compares the LES and DNS simulations in terms of efficiency, and fraction of power derived from the angular motion of the foil ($C_{p,ang} / C_{p,tot}$) for 15 kinematics across two frequencies.

\begin{table}[!htbp]
\centering
\caption{Efficiency $\eta$ of DNS simulations at $Re=1000$ compared with LES at $Re=50,000$. $\Delta \eta=\eta_{LES}-\eta_{DNS}$, and $C_{p,ang} / C_{p,tot}$ is the fraction of power derived from the angular motion of the foil.}
\label{tab:table2}
\setlength{\tabcolsep}{6pt}
\begin{tabular}{lccccc}
\hline \hline
\multicolumn{1}{l}{Kinematics} & \multicolumn{1}{c}{$\eta$} & \multicolumn{1}{c}{$\eta$} &
\multicolumn{1}{c}{$\Delta \eta$} & 
\multicolumn{1}{c}{${C}_{p,ang}/{C}_{p,tot}$} & 
\multicolumn{1}{c}{${C}_{p,ang}/{C}_{p,tot}$} \\
              & \multicolumn{1}{c}{{DNS}} & \multicolumn{1}{c}{{LES}} &  &
\multicolumn{1}{c}{{DNS}} & 
\multicolumn{1}{c}{{LES}} \\ \hline
\textbf{$\mathbf {fc/U_{\infty}=0.1}$}          &                                &                       &                                           &                 &                   \\ \hline
$h_o/c=0.5$ $\theta_o$=$60\,^{\circ}$ & \multicolumn{1}{c}{0.240}   & \multicolumn{1}{c}{0.268}    & \multicolumn{1}{c}{0.028}             & \multicolumn{1}{c}{0.13}        & \multicolumn{1}{c}{0.12}         \\ 
$h_o/c=1$ $\theta_o$=$65\,^{\circ}$   & \multicolumn{1}{c}{0.256}   & \multicolumn{1}{c}{0.281}    & \multicolumn{1}{c}{0.025}           & \multicolumn{1}{c}{0.06}         & \multicolumn{1}{c}{0.05}         \\ 
$h_o/c=1.5$ $\theta_o$=$65\,^{\circ}$ & \multicolumn{1}{c}{0.208}   & \multicolumn{1}{c}{0.222}    & \multicolumn{1}{c}{0.014}       & \multicolumn{1}{c}{-0.02}        & \multicolumn{1}{c}{-0.03}       \\ 
$h_o/c=0.5$ $\theta_o$=$75\,^{\circ}$ & \multicolumn{1}{c}{0.201}   & \multicolumn{1}{c}{0.226}    & \multicolumn{1}{c}{0.025}        & \multicolumn{1}{c}{0.21}        & \multicolumn{1}{c}{0.22}      \\ 
$h_o/c=1$ $\theta_o$=$75\,^{\circ}$   & \multicolumn{1}{c}{0.230}   & \multicolumn{1}{c}{0.264}    & \multicolumn{1}{c}{0.034}          & \multicolumn{1}{c}{0.10}        & \multicolumn{1}{c}{0.10}          \\ 
$h_o/c=2$ $\theta_o$=$75\,^{\circ}$   & \multicolumn{1}{c}{0.222}   & \multicolumn{1}{c}{0.246}    & \multicolumn{1}{c}{0.024}         & \multicolumn{1}{c}{-0.01}        & \multicolumn{1}{c}{-0.01}       \\ 
$h_o/c=1$ $\theta_o$=$85\,^{\circ}$   & \multicolumn{1}{c}{0.181}   & \multicolumn{1}{c}{0.217}    & \multicolumn{1}{c}{0.036}          & \multicolumn{1}{c}{0.16}        & \multicolumn{1}{c}{0.17}        \\ 
$h_o/c=2$ $\theta_o$=$85\,^{\circ}$   & \multicolumn{1}{c}{0.187}   & \multicolumn{1}{c}{0.223}    & \multicolumn{1}{c}{0.036}         & \multicolumn{1}{c}{0.04}         & \multicolumn{1}{c}{0.05}         \\ \hline
\textbf{$\mathbf {fc/U_{\infty}=0.15}$}         &                                &                                 &                                           &                                  &	                    \\ \hline
$h_o/c=1$ $\theta_o$=$65\,^{\circ}$   & \multicolumn{1}{c}{0.200}   & \multicolumn{1}{c}{0.250}    & \multicolumn{1}{c}{0.050}         & \multicolumn{1}{c}{-0.18}       & \multicolumn{1}{c}{-0.16}        \\ 
$h_o/c=1$ $\theta_o$=$75\,^{\circ}$   & \multicolumn{1}{c}{0.255}   & \multicolumn{1}{c}{0.300}    & \multicolumn{1}{c}{0.045}         & \multicolumn{1}{c}{-0.10}       & \multicolumn{1}{c}{-0.12}       \\ 
$h_o/c=1.5$ $\theta_o$=$75\,^{\circ}$ & \multicolumn{1}{c}{0.183}   & \multicolumn{1}{c}{0.235}    & \multicolumn{1}{c}{0.052}       & \multicolumn{1}{c}{-0.23}       & \multicolumn{1}{c}{-0.22}          \\ 
$h_o/c=2$ $\theta_o$=$75\,^{\circ}$   & \multicolumn{1}{c}{0.072}    & \multicolumn{1}{c}{0.139}    & \multicolumn{1}{c}{0.067}         & \multicolumn{1}{c}{-0.58}       & \multicolumn{1}{c}{-0.34}      \\ 
$h_o/c=1$ $\theta_o$=$85\,^{\circ}$   & \multicolumn{1}{c}{0.266}   & \multicolumn{1}{c}{0.274}    & \multicolumn{1}{c}{0.008}          & \multicolumn{1}{c}{0.02}         & \multicolumn{1}{c}{-0.08}       \\ 
$h_o/c=1.5$ $\theta_o$=$85\,^{\circ}$ & \multicolumn{1}{c}{0.222}   & \multicolumn{1}{c}{0.284}    & \multicolumn{1}{c}{0.062}       & \multicolumn{1}{c}{-0.16}       & \multicolumn{1}{c}{-0.11}        \\ 
$h_o/c=2$ $\theta_o$=$85\,^{\circ}$   & \multicolumn{1}{c}{0.179}   & \multicolumn{1}{c}{0.206}    & \multicolumn{1}{c}{0.027}         & \multicolumn{1}{c}{-0.28}       & \multicolumn{1}{c}{-0.22}        \\ \hline \hline
\end{tabular}
\end{table}

Although the general trends are consistent across the Reynolds number regimes investigated, results in Table \ref{tab:table2} demonstrate that there is always a small to moderate increase in efficiency ($\Delta \eta$) with the higher Reynolds number. These ranged from $\Delta \eta = 1.4 - 3.6\%$ for kinematics at $fc/U_{\infty}=0.1$ and $\Delta \eta = 0.8 - 6.7\%$ for kinematics at $fc/U_{\infty}=0.15$.

Although the translational motion consistently contributed the majority of the power, the role of the angular power varies significantly among the high-efficiency kinematics explored in Table \ref{tab:table2}. For kinematics at $fc/U_{\infty}=0.1$, the translational and angular power increase proportionally with an increase in Reynolds number, roughly maintaining the percent of total power that comes from angular power. Most of these cases have a negligible or positive contribution. As the frequency increases to $fc/U_{\infty}=0.15$ the contribution is now negative for both Reynolds number regimes, but the ratio varies significantly from $-0.58$ to $+0.02$. The largest contributor of negative power at $C_{p,ang} / C_{p,tot}=-0.58$ is for $h_o/c=2$ and $\theta_o=75^{\circ}$, in which the total efficiency is significantly impacted resulting in $\eta=0.072$ for the DNS.  The high Reynolds number simulation for these kinematics only slightly increased the efficiency ($\eta=0.139$) due to less negative angular power contributions. The opposite extreme is when the angular power contributions are close to zero, for example the kinematics of $fc/U_{\infty}=0.15$, $h_o/c=1$ and $\theta_o=85^{\circ}$, which have efficiency values of $\eta=0.266$ and $\eta=0.274$ for low and high Reynolds number respectively.
\subsection{Effects of Reynolds number}

\begin{figure}[!htbp]
\centering
	\begin{subfigure}{0.49\textwidth}
      	\includegraphics[width=1\linewidth]{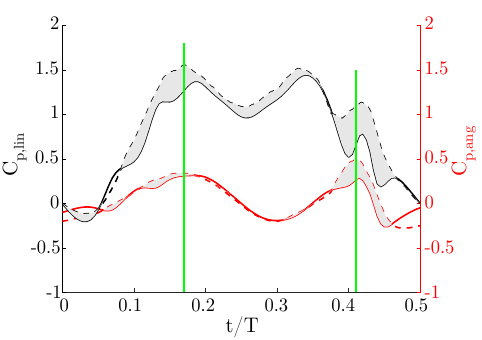}
		\caption{Linear ($C_{p,lin}$), angular power ($C_{p,ang}$)}
		\label{fig:f1power}
	\end{subfigure}
	\begin{subfigure}{0.49\textwidth}
		\includegraphics[width=1\linewidth]{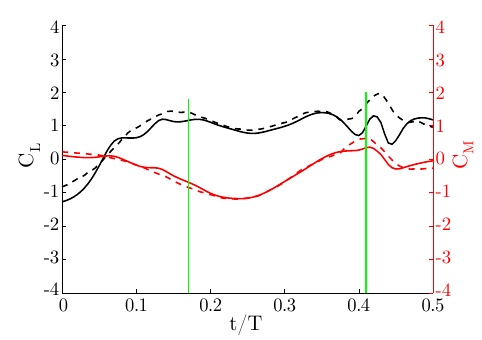}
		\caption{Lift forces ($C_L$) and moment ($C_M$)}
		\label{fig:f1lift}
	\end{subfigure}
    \begin{subfigure}{0.5\textwidth}
      	\makebox[\textwidth][c]{\includegraphics[width=5.5cm]{Flowfields_legend.pdf}}
    \end{subfigure}
    
	\begin{subfigure}{0.49\textwidth}
		\includegraphics[width=0.49\textwidth]{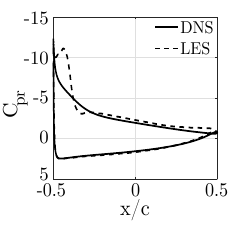}
		\includegraphics[width=0.49\textwidth]{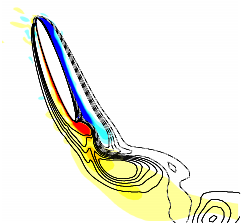}
		\caption{t/T = 0.17}
		\label{fig:f1Cpflow1}
	\end{subfigure}
	\begin{subfigure}{0.49\textwidth}
		\includegraphics[width=0.49\textwidth]{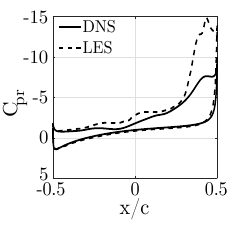}
		\includegraphics[width=0.49\textwidth]{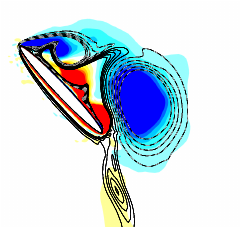}
		\caption{t/T = 0.41}
		\label{fig:f1Cpflow2}
	\end{subfigure}
\caption{Direct comparison of the phase-averaged power, lift force, torque and pressure coefficient between DNS (---) and LES (- - -) during an upstroke. The gray regions represent the times when LES achieved higher power than DNS. Kinematics: $fc/U_{\infty}=0.1$, $h_o/c=2$, $\theta_o=85^{\circ}$. The phase and span-averaged vorticity fields in (c) and (d) are represented by the green lines in (b), where colors are LES and lines are DNS and the dashed lines represent negative vorticity, solid lines represent positive vorticity.}
\label{fig:forcesf1}
\end{figure}

\begin{figure}[!htbp]
\centering
	\begin{subfigure}{0.49\textwidth}
      	\includegraphics[width=1\linewidth]{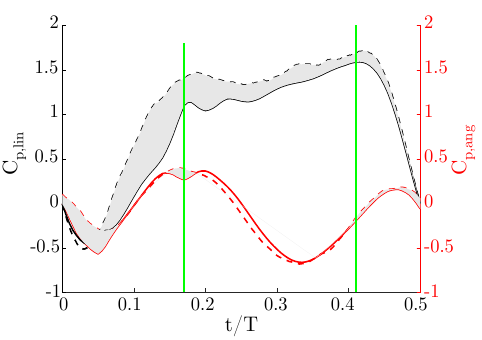}
		\caption{Linear ($C_{p,lin}$), angular power ($C_{p,ang}$)}
		\label{fig:f15power}
	\end{subfigure}
	\begin{subfigure}{0.49\textwidth}
		\includegraphics[width=1\linewidth]{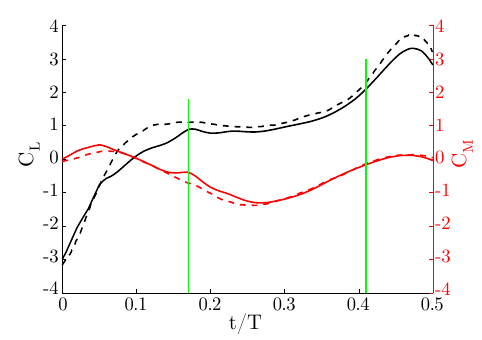}
		\caption{Lift forces ($C_L$) and moment ($C_M$)}
		\label{fig:f15lift}
	\end{subfigure}
    \begin{subfigure}{0.5\textwidth}
      	\makebox[\textwidth][c]{\includegraphics[width=5.5cm]{Flowfields_legend.pdf}}
    \end{subfigure}
    
	\begin{subfigure}{0.49\textwidth}
		\includegraphics[width=0.49\textwidth]{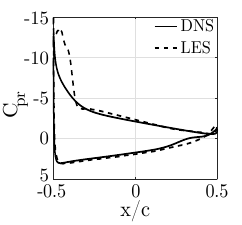}
		\includegraphics[width=0.49\textwidth]{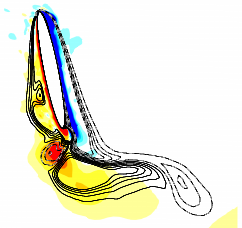}
		\caption{t/T = 0.17}
		\label{fig:f15Cpflow1}
	\end{subfigure}
	\begin{subfigure}{0.49\textwidth}
		\includegraphics[width=0.49\textwidth]{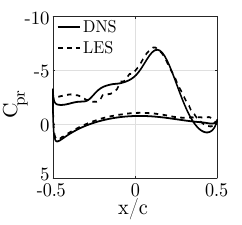}
		\includegraphics[width=0.49\textwidth]{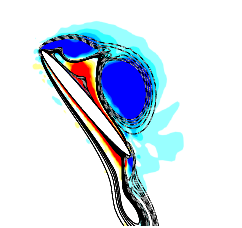}
		\caption{t/T = 0.41}
		\label{fig:f15Cpflow2}
	\end{subfigure}
\caption{See caption from Fig. \ref{fig:forcesf1}. Kinematics: $fc/U_{\infty}=0.15$,  $h_o/c=1.5$, $\theta_o=85^{\circ}$.}
\label{fig:forces}
\end{figure}

To show a more thorough comparison across the two frequency and two Reynolds number regimes, Figs. \ref{fig:forcesf1} and \ref{fig:forces} show the time-dependent power, forces and moments for two sets of kinematics for the upstroke, from $t/T=0-0.5$. The green lines in both figures are located at $t/T=0.17$ and $t/T=0.41$ and their respective pressure coefficients ($C_{pr}$) on foil surfaces and vorticity fields can be seen in Figs. \ref{fig:f1Cpflow1}, \ref{fig:f1Cpflow2}, \ref{fig:f15Cpflow1}, \ref{fig:f15Cpflow2}.

In Fig. \ref{fig:forcesf1} at the lower reduced frequency of $fc/U_{\infty}=0.1$, the heave and pitch amplitudes are $h_o/c=2$ and $\theta_o=85^{\circ}$. As shown by the LES, the higher Reynolds number produces an increase in power of $\Delta \eta=0.036$, with a small positive contribution from the angular power of $4-5\%$. The higher Reynolds number flow produces more translational power throughout the stroke compared with the low Reynolds number flow. Shown in Fig. \ref{fig:f1Cpflow1} at $t/T=0.17$, a LEV has begun to form at the leading edge, which is also shown in the $C_{pr}$ peak, enhancing the lift force. This enhancement occurs during a very high vertical velocity amplifying power production. The translational power peaks again around $t/T=0.35$ corresponding with the formation of a second LEV. The increase in both the translational and angular contribution of power is prominent at $t/T=0.41$, as the primary LEV begins to shed and a secondary LEV is providing lift enhancement. At this point an opposite sign trailing edge vortex (TEV) has developed in the high Reynolds number flow, whereas the low Reynolds number flow maintains a shear layer of positive vorticity on the suction side of the foil. The TEV increases the pressure gradient corresponding to the large LES pressure coefficient on the trailing edge, and contributes positively to translational and angular power generation.

Fig. \ref{fig:forces} shows the direct comparison between low and high Reynolds number at the higher frequency of $fc/U_{\infty}=0.15$, with heave and pitch amplitudes of $h_o/c=1.5$ and $\theta_o=85^{\circ}$. This set of kinematics at higher reduced frequency shows a significant increase in translational power over the first half of the upstroke when the LEV is developing and increasing the lift force, whereas the angular power does not show much discrepancy. The increase in lift force results in an efficiency increase of $\Delta \eta=0.062$. For the higher frequency early in the upstroke in Fig. \ref{fig:f15Cpflow1}, the vorticity field has a secondary counter-clockwise rotating LEV from the previous downstroke that has not yet been shed from the pressure side of the foil. By $t/T=0.41$ these vortices have shed but a new set of two clockwise rotating LEVs are present on the suction surface. The faster stroke kinematics at $fc/U_{\infty}=0.15$ give the LEV less time to shed and convect downstream, producing LEVs that are closer to the foil surface and thus remain on the foil for a higher percentage of the stroke. 

\subsection{Vortex wake dynamics and trajectory} \label{wakedynamics}

The location and strength of one or more LEVs can significantly modify the energy harvesting potential of a foil. Furthermore, once the LEV is shed, its trajectory can also impact the performance of downstream foils. To better understand the LEV formation and resulting trajectory, this section tracks the location of the primary LEV formation on the foil and in the near wake region for a subset of high-efficiency kinematics.

The tracking algorithm relies on a user input to select the LEV to be tracked shortly after it is formed. The primary LEV, or the first LEV that rolls up from the leading edge on the upstroke will be tracked. At every $tU_{\infty}/c=0.1$ timesteps the position of the maximum value of vorticity for this specific vortex is tracked as it forms on the foil and then sheds downstream for approximately three chord lengths. In order to clearly identify the primary LEV the $fc/U_{\infty}=0.1$ kinematics have $t/T = 0.25$ as an initial time and the $fc/U_{\infty}=0.15$ kinematics start tracking at $t/T = 0.30$ due to the delayed LEV formation.

\begin{figure}[!htbp]
\centering
	    \includegraphics[width=0.5\textwidth]{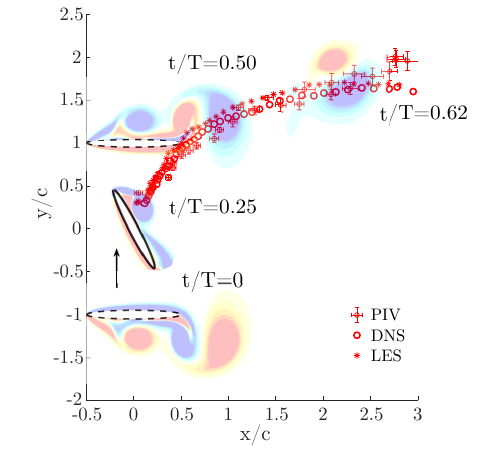}
		\caption{LEV trajectory from PIV, DNS and LES superimposed with contours from DNS to illustrate LEV size and position at t/T = 0, 0.25, 0.50 and 0.62. Kinematics: $fc/U_{\infty} = 0.1, h_o/c=1, \theta_o=65^{\circ}$.}
		\label{fig:validation}
\end{figure}

Fig. \ref{fig:validation} shows the vortex tracking from phase-averaged PIV \citep{SuThesis}, DNS and LES flow fields for the kinematics $fc/U_{\infty} = 0.1$, $h_o/c=1$ and $\theta_o=65^{\circ}$. Overlaid with the primary LEV path is the foil motion during its upstroke and the vorticity fields from DNS to show the LEV size and position relative to the trajectory at $t/T=0$, $0.25$, $0.50$, and $0.62$. There is strong agreement between PIV, DNS and LES, in terms of vortex position and shedding time. The LEV decays faster with increasing Reynolds number, and the vortices become more difficult to track as shown by the PIV data at approximately $x/c=2.5$ where the tracking algorithm has a larger discrepancy shown by the increased size of the PIV error bars in Fig. \ref{fig:validation}. The PIV data is more difficult to accurately track far from the foil, which is likely due to the three-dimensional wingtip effects in the experiments compared with the infinite-span model in the DNS and LES. Due to the similarity between the two sets of simulations and experiments, only DNS tracking is discussed in the following analysis.

\begin{figure}[H]
        \centering
        \begin{subfigure}{0.49\textwidth}
      		\includegraphics[width=1\textwidth]{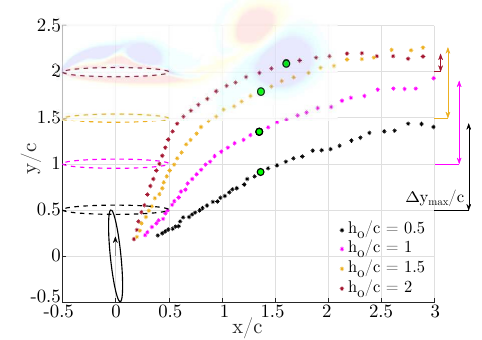}
      		\caption{$fc/U_{\infty} = 0.1$ and $\theta_o = 85^{\circ}$}
      		\label{fig:changeheave}
	    \end{subfigure}
	    \begin{subfigure}{0.49\textwidth}
      		\includegraphics[width=1\textwidth]{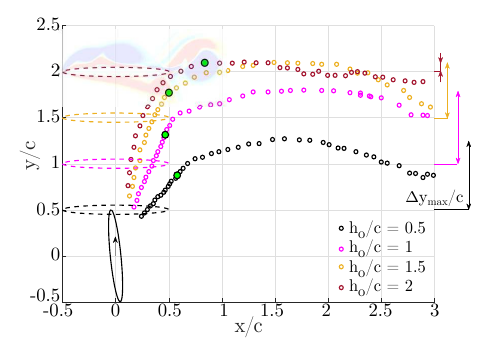}
      		\caption{$fc/U_{\infty} = 0.15$ and $\theta_o = 85^{\circ}$}
      		\label{fig:changefreq}
	    \end{subfigure}
	\caption{Effect of heave amplitude, $h_o/c$, on LEV trajectory. Contours at the top of upstroke for $h_o/c = 2$ illustrate differences of the LEV core location in two frequencies with fixed $\theta_o = 85^{\circ}$.}
\label{fig:heaveandfreq}
\end{figure}

The path of the LEV varies significantly with changing kinematics. Fig. \ref{fig:heaveandfreq} shows the effect of heave amplitude on the LEV dynamics for fixed pitch amplitude $\theta_{o}=85^{\circ}$ and two reduced frequencies, $fc/U_{\infty}=0.1$ and $0.15$. The solid line is the foil position at mid-upstroke, $t/T = 0.25$, and the path of the LEV corresponds to the four colored foils representing heave amplitudes $h_o/c=0.5$, $1$, $1.5$, and $2$. With increasing heave amplitude, the maximum vertical distance traveled by the LEV saturates. In Fig. \ref{fig:changeheave}, comparing the vertical distance with the top of the heave stroke, the value of $\Delta y_{max}/c$ is almost zero at $h_o/c=2$, meaning the LEV maintains the same vertical position as the foil. In comparison at low heave of $h_o/c=0.5$, the LEV travels approximately 1 chord length higher than the maximum position of the foil, or $\Delta y_{max}/c=1$. This behavior is similar at $fc/U_{\infty}=0.15$ in Fig. \ref{fig:changefreq}, except $\Delta y_{max}/c\approx0.75$ at its maximum. Thus, a nonlinear dependence of $\Delta y_{max}/c$ with heave amplitude is found in these two frequencies.

In the background of Figs. \ref{fig:changeheave} and \ref{fig:changefreq} the vorticity field at $t/T = 0.50$ shows the LEV location for the heave amplitude $h_o/c = 2$, and the LEV core position is given by the green marker. This marker of $t/T=0.5$ is represented in the lower heave amplitudes as well.  The horizontal position of the vortex is very similar for $h_o/c=0.5-1.5$, however at $h_o/c=2$ it has traveled further downstream. This is due to an earlier shedding time at this high amplitude, and the LEV separates from the foil prior to completing the upstroke. This in turn allows the vortex to travel more horizontal distance but limits its vertical motion. 

The frequency also significantly alters the path of the primary LEV between Figs. \ref{fig:changeheave} and \ref{fig:changefreq}. At $fc/U_{\infty}=0.15$ the trajectory is linear while the LEV is close to the foil then turns abruptly in a horizontal trajectory downstream.  During the relatively fast heaving motion the LEV convects with the foil in a rigid body motion, hence limiting the LEV size and horizontal location. This also explains the near-foil location of the LEV at $t/T=0.5$ indicated by the green markers in Fig. \ref{fig:changefreq}. At $fc/U_{\infty}=0.1$ the stroke is relatively slower, the LEV grows larger and is shed earlier in the cycle, and travels further downstream by $t/T=0.5$, as indicated by the green markers in Fig. \ref{fig:changeheave}. 

\begin{figure}[H]
        \centering
            \includegraphics[width=0.5\textwidth]{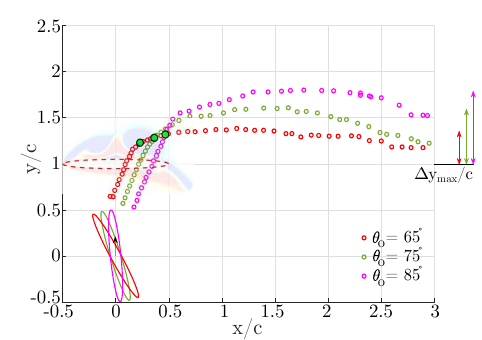}
	\caption{Effect of pitch amplitude, $\theta_o$, on LEV trajectory with $fc/U_{\infty} = 0.15$ held constant. Contours at the top of upstroke for $h_o/c = 1$ illustrate the LEV core location for $\theta_o = 65^{\circ}$.}
\label{fig:changeaoa}
\end{figure}

In a similar analysis the frequency and heave are held constant ($fc/U_{\infty} = 0.15$ and $h_o/c = 1$) and the effect of pitch amplitude on the LEV trajectory is shown in Fig. \ref{fig:changeaoa}. As the pitch amplitude increases the LEV grows to a larger size and is located closer to the trailing edge at stroke reversal (green markers) due to the larger angular velocity within an equal heaving distance. Within the high-efficiency energy harvesting range that is explored between $\theta_o=65^{\circ}$ to $85^{\circ}$, the LEV size and maximum height follows a linear dependence with pitch amplitude.

\section{Conclusion}
The energy harvesting mechanisms of an elliptical oscillating foil are explored for two non-dimensional frequencies, $fc/U_{\infty}=0.1$ and $0.15$ at $Re=1000$ and $Re=50,000$ with the goal of exploring the differences in power generation and vortex dynamics across the two Reynolds numbers. A sweep of pitch and heave amplitudes were performed at $Re=1000$ and those kinematics that yielded the highest efficiency or power generation capability were repeated with LES at $Re=50,000$, and compared with available experimental results. It is found that there were only minor variations in the energy extraction and vortex dynamics between these two Reynolds number regimes.  For the lower frequency of $fc/U_{\infty}=0.1$ the maximum efficiency of $25.6\%$ occurs at a heave amplitude of $h_o/c=1$ and pitch amplitude $\theta_o=65^{\circ}$, and is increased to $28.1\%$ for a Reynolds number of $50,000$. This modest increase in power production is seen with all the high Reynolds number cases at $fc/U_{\infty}=0.1$.  At the higher frequency of oscillation, $fc/U_{\infty}=0.15$, the high and low Reynolds number results show more variability, with the $Re=50,000$ flow extracting $0.8-6.7\%$ more energy than the same kinematics at $Re=1000$. 

A persistent feature of all high-efficiency kinematics is a coherent LEV that forms on the heave stroke, and sheds at or just after the heave stroke reversal. This occurs at high relative angles of attack, which is consistent with large heave and/or pitch amplitudes. Comparing all data in terms of relative angle of attack at mid-upstroke, or $\alpha_{T/4}$, it is shown that the maximum efficiency for each heave occurs when $\alpha_{T/4}>22^{\circ}$, or when the flow has clearly separated from the foil. As the flow separates at the leading edge it forms a LEV that persists throughout the upstroke, enhancing the lift, and thus the power extraction from the translational motion. When comparing the two Reynolds number regimes, the increase in efficiency with higher Reynolds number is from a slightly stronger LEV that forms earlier in the heave cycle, and enhances the power extraction from both lift and torque during critical portions of the cycle.

The details of this LEV formation, shedding, and trajectory are examined through tracking its core position through phase-averaged data. In comparing experiments with DNS and LES results, it is found that the primary LEV position downstream is relatively independent of Reynolds number, but that it is strongly influenced by its kinematics. The maximum vertical distance traveled by the primary LEV is reported, as well as its trajectory up to 3 chord lengths downstream. As the heave amplitude is increased to $2c$ the maximum vertical distance saturates because the LEV is shed earlier in the heave process, resulting in a relatively horizontal trajectory. At a reduced frequency of $fc/U_{\infty}=0.1$ the primary LEV is almost one full chord length downstream by mid-stroke, and the vertical trajectory dependent on pitch and heave amplitude. As the reduced frequency is increased to $fc/U_{\infty}=0.15$ the LEV has less time to develop, resulting in the primary LEV very close to the trailing edge at mid-stroke.

The implications of the Reynolds number independence in terms of efficiency and vortex dynamics are important to understand how oscillating foil energy harvesting devices will scale with size and flow speed. Furthermore the impact of the coherent vortices in the wake is likely to be an important factor in array design and configuration.

\section{Acknowledgments}

Funding for this research has come from the Advanced Research Projects Agency-Energy, United States of America (ARPA-E award DE-AR0000318), National Science Foundation, United States of America (CBET award 1921594), and a Karen T. Romer Undergraduate Teaching and Research Award, Brown University. The authors thank Kenny Breuer, Yunxing Su and Michael J. Miller for access to their experimental data. This research was conducted using computational resources and services at the Center for Computation and Visualization, Brown University.

\section{References}

\bibliography{paper_reviewed_2019}

\begin{thebibliography}{10}
\expandafter\ifx\csname url\endcsname\relax
  \def\url#1{\texttt{#1}}\fi
\expandafter\ifx\csname urlprefix\endcsname\relax\def\urlprefix{URL }\fi
\expandafter\ifx\csname href\endcsname\relax
  \def\href#1#2{#2} \def\path#1{#1}\fi

\bibitem{doe}
{Wind and Water Power Program},
  \href{http://www1.eere.energy.gov/water/pdfs/mhk_project_2011.pdf}{Marine and
  {H}ydrokinetic {E}nergy {P}rojects}, U.S. Department of Energy-Office of
  Energy Efficiency and Renewable Energy, 2012, {T}echnical Report DOE/EE-0710.
\newline\urlprefix\url{http://www1.eere.energy.gov/water/pdfs/mhk_project_2011.pdf}

\bibitem{gorlov98}
A.~Gorlov,
  \href{https://www.osti.gov/scitech/servlets/purl/666280/}{Development of the
  {H}elical {R}eaction {H}ydraulic {T}urbine}, U.S. Department of Energy, 1998,
  {T}echnical Report DOE/EE/15669-T1.
\newline\urlprefix\url{https://www.osti.gov/scitech/servlets/purl/666280/}

\bibitem{mckdel}
W.~McKinney, J.~DeLaurier, Wingmill: An oscillating-wing windmill, J. Energy
  5~(2) (1981) 109--115.
\newblock \href {http://dx.doi.org/10.2514/3.62510}
  {\path{doi:10.2514/3.62510}}.

\bibitem{Young2014}
J.~Young, J.~Lai, M.~Platzer, A review of progress and challenges in flapping
  foil power generation, Prog. Aerosp. Sci. 67~(1) (2014) 2--28.
\newblock \href {http://dx.doi.org/10.1016/j.paerosci.2013.11.001}
  {\path{doi:10.1016/j.paerosci.2013.11.001}}.

\bibitem{Xiao2014}
Q.~Xiao, Q.~Zhu, A review on flow energy harvesters based on flapping foils, J.
  Fluids Struct. 46 (2014) 174--191.
\newblock \href {http://dx.doi.org/10.1016/j.jfluidstructs.2014.01.002}
  {\path{doi:10.1016/j.jfluidstructs.2014.01.002}}.

\bibitem{liu2013}
W.~Liu, Q.~Xiao, F.~Cheng, A bio-inspired study on tidal energy extraction with
  flexible flapping wings, Bioinspir. Biomim. 8~(3) (2013) .
\newblock \href {http://dx.doi.org/10.1088/1748-3182/8/3/036011}
  {\path{doi:10.1088/1748-3182/8/3/036011}}.

\bibitem{SuThesis}
Y.~Su, Energy {H}arvesting and {A}eroelastic {I}nstabilities {U}sing
  {P}rescribed and {E}lastically-{M}ounted {P}itching and {H}eaving
  {H}ydrofoils, ({Ph.D.} thesis), Brown University (2019).

\bibitem{simpson2008}
B.~J. Simpson, S.~Licht, F.~S. Hover, M.~S. Triantafyllou, Energy extraction
  through flapping foils, {ASME} 27th International Conference on Offshore
  Mechanics and Arctic Engineering. American Society of Mechanical Engineers,
  Estoril, Portugal, 2008, pp. 389--395.
\newblock \href {http://dx.doi.org/10.1115/OMAE2008-58043}
  {\path{doi:10.1115/OMAE2008-58043}}.

\bibitem{Kim2017}
D.~Kim, B.~Strom, S.~Mandre, K.~S. Breuer, Energy harvesting performance and
  flow structure of an oscillating hydrofoil with finite span, J. Fluids
  Struct. 70 (2017) 314--326.
\newblock \href {http://dx.doi.org/10.1016/j.jfluidstructs.2017.02.004}
  {\path{doi:10.1016/j.jfluidstructs.2017.02.004}}.

\bibitem{Yunxing2019}
Y.~Su, M.~Miller, S.~Mandre, K.~Breuer, Confinement effects on energy
  harvesting by a heaving and pitching hydrofoil, J. Fluids Struct. 84 (2019)
  233--242.
\newblock \href {http://dx.doi.org/10.1016/j.jfluidstructs.2018.11.006}
  {\path{doi:10.1016/j.jfluidstructs.2018.11.006}}.

\bibitem{kindum2011}
T.~Kinsey, G.~Dumas, G.~Lalande, J.~Ruel, A.~Méhut, P.~Viarouge, J.~Lemay,
  Y.~Jean, Prototype testing of a hydrokinetic turbine based on oscillating
  hydrofoils, Renew. Energy 36~(6) (2011) 1710--1718.
\newblock \href {http://dx.doi.org/10.1016/j.renene.2010.11.037}
  {\path{doi:10.1016/j.renene.2010.11.037}}.

\bibitem{cardona2015}
J.~L. Cardona, M.~J. Miller, T.~Derecktor, S.~Winckler, K.~Volkmann, A.~Medina,
  S.~Cowles, R.~Lorick, K.~S. Breuer, S.~Mandre, Field testing of a 1 kw
  oscillating hydrofoil energy harvesting system, Marine Energy Technology
  Symposium, Washington D.C., USA, 2016.

\bibitem{MillerThesis}
M.~Miller, Fluid {D}ynamic {R}esearch and {P}rototype {D}evelopment of the
  {L}eading {E}dge {O}scillating {H}ydrofoil for {H}ydrokinetic {E}nergy
  {H}arvesting, ({Ph.D.} thesis), Brown University (2019).

\bibitem{kindum2008}
T.~Kinsey, G.~Dumas, Parametric study of an oscillating airfoil in a
  power-extraction regime, AIAA J. 46~(6) (2008) 1318--1330.
\newblock \href {http://dx.doi.org/10.2514/1.26253}
  {\path{doi:10.2514/1.26253}}.

\bibitem{zhu2011}
Q.~Zhu, Optimal frequency for flow energy harvesting of a flapping foil, J.
  Fluid Mech. 675 (2011) 495--517.
\newblock \href {http://dx.doi.org/10.1017/S0022112011000334}
  {\path{doi:10.1017/S0022112011000334}}.

\bibitem{wu2015}
J.~Wu, Y.~L. Chen, N.~Zhao, Role of induced vortex interaction in a semi-active
  flapping foil based energy harvester, Phys. Fluids 27~(9) (2015) .
\newblock \href {http://dx.doi.org/10.1063/1.4930028}
  {\path{doi:10.1063/1.4930028}}.

\bibitem{Su2019}
Y.~Su, K.~Breuer, Resonant response and optimal energy harvesting of an
  elastically mounted pitching and heaving hydrofoil, Phys. Rev. Fluids 4~(6)
  (2019) .
\newblock \href {http://dx.doi.org/10.1103/PhysRevFluids.4.064701}
  {\path{doi:10.1103/PhysRevFluids.4.064701}}.

\bibitem{baik2012}
Y.~S. Baik, L.~P. Bernal, K.~Granlund, M.~V. Ol, Unsteady force generation and
  vortex dynamics of pitching and plunging aerofoils, J. Fluid Mech. 709 (2012)
  37--68.
\newblock \href {http://dx.doi.org/10.1017/jfm.2012.318}
  {\path{doi:10.1017/jfm.2012.318}}.

\bibitem{Bernitsas2014}
L.~Ding, L.~Zhang, M.~M. Bernitsas, C.-C. Chang, Numerical simulation and
  experimental validation for energy harvesting of single-cylinder vivace
  converter with passive turbulence control, Renew. Energy J. 85 (2015)
  1246--1259.
\newblock \href {http://dx.doi.org/10.1016/j.renene.2015.07.088}
  {\path{doi:10.1016/j.renene.2015.07.088}}.

\bibitem{Simeski2017}
F.~Simeski, Simulations of {H}ydrofoil {A}rrays with {A}pplications in {E}nergy
  {H}arvesting, ({B}achelor's thesis), Brown University (2017).

\bibitem{Ashraf2011}
M.~A. Ashraf, J.~Young, J.~C.~S. Lai, M.~F. Platzer, Numerical analysis of an
  oscillating-wing wind and hydropower generator, AIAA J. 49~(7) (2011)
  1374--1386.
\newblock \href {http://dx.doi.org/10.2514/1.J050577}
  {\path{doi:10.2514/1.J050577}}.

\bibitem{Xiao2012}
Q.~Xiao, W.~Liao, S.~Yang, Y.~Peng, How motion trajectory affects energy
  extraction performance of a biomimic energy generator with an oscillating
  foil?, Renew. Energy 37~(1) (2012) 61--75.
\newblock \href {http://dx.doi.org/10.1016/j.renene.2011.05.029}
  {\path{doi:10.1016/j.renene.2011.05.029}}.

\bibitem{kinsey2012}
T.~Kinsey, G.~Dumas, Computational fluid dynamics analysis of a hydrokinetic
  turbine based on oscillating hydrofoils, J. Fluids Eng. 134~(2) (2012) .
\newblock \href {http://dx.doi.org/10.1115/1.4005841}
  {\path{doi:10.1115/1.4005841}}.

\bibitem{campobasso2013}
M.~S. Campobasso, A.~Piskopakis, M.~Yan, Analysis of an oscillating wing in a
  power-extraction regime based on the compressible reynolds-averaged
  navier-stokes equations and the k--$\omega$ {SST} turbulence model, {ASME}
  Turbo Expo: Turbine Technical Conference and Exposition. American Society of
  Mechanical Engineers, San Antonio, USA, 2013.
\newblock \href {http://dx.doi.org/10.1115/GT2013-94531}
  {\path{doi:10.1115/GT2013-94531}}.

\bibitem{franck2010}
J.~A. Franck, T.~Colonius, A compressible large-{E}ddy simulation of separation
  control on a wall-mounted hump, AIAA J. 48~(6) (2010) 1098--1107.
\newblock \href {http://dx.doi.org/10.2514/1.44756}
  {\path{doi:10.2514/1.44756}}.

\bibitem{Greenblatt2004}
D.~Greenblatt, K.~B. Paschal, C.-S. Yao, J.~Harris, N.~W. Schaeffler, A.~E.
  Washburn, A separation control {CFD} validation test case, {P}art 1: baseline
  and steady suction, AIAA J. 2220 (2004) .
\newblock \href {http://dx.doi.org/10.2514/6.2004-2220}
  {\path{doi:10.2514/6.2004-2220}}.

\end{thebibliography}

\end{document}